\documentclass[aps,groupedaddress,nofootinbib,showkeys,showpacs,amsfonts,eqsecnum]{revtex4}

\usepackage{bm}
\usepackage{amsmath}

\newcommand{\R}{{\bm {R}}}
\newcommand{\Rc}{{\cal {R}}}
\newcommand{\tr}{{\textrm {tr}}}
\newcommand{\Tr}{{\textrm {Tr}}}
\newcommand{\Det}{{\textrm {Det}}}

\begin{document}

\title{The method of covariant symbols in curved space-time}

\author{L.L. Salcedo}
\email{salcedo@ugr.es}

\affiliation{
Departamento de F\'{\i}sica At\'omica, Molecular y Nuclear, Universidad de Granada, E-18071
Granada, Spain }

\date{\today}

\begin{abstract}
Diagonal matrix elements of pseudodifferential operators are needed in
order to compute effective Lagrangians and currents. For this purpose
the method of symbols is often used, which however lacks manifest
covariance.  In this work the method of covariant symbols, introduced
by Pletnev and Banin, is extended to curved space-time with arbitrary
gauge and coordinate connections. For the Riemannian connection we
compute the covariant symbols corresponding to external fields, the
covariant derivative and the Laplacian, to fourth order in a covariant
derivative expansion. This allows to obtain the covariant symbol of
general operators to the same order. The procedure is illustrated by
computing the diagonal matrix element of a nontrivial operator to
second order. Applications of the method are discussed.
\end{abstract}



\maketitle

\section{Introduction}
\label{sec:1}

As is known, the functional integral formulation of quantum field
theory depends on the computation of the partition functional. To one
loop this amounts to adding the quadratic fluctuations above a
classical solution. Typically
\begin{equation}
Z_{{\text{1-loop}}}= \int D{\psi}(x)
\, e^{-\langle \psi|{\hat K}|\psi\rangle }
\end{equation}
where the quantum fluctuations are controlled by the differential
operator $\hat K$. This operator may depend on all kind of external
fields and typically it will contain the covariant derivative
$\nabla_\mu$ (with all kind of connections) as well as other background
fields, $M(x)$, i.e. $\hat K= K(\nabla,M)$. Formally, the Gaussian
integral gives the functional determinant $\Det(\hat K)$ raised to
some power which depends on the type of fields (real or complex,
bosonic or fermionic). Thus, for the effective action, 
\begin{equation}
Z= e^{-\Gamma}\,,
\end{equation}
one formally obtains
\begin{equation}
\Gamma_{{\text{1-loop}}}= c\,\Tr\log(\hat K) 
=c\int d^dx\sqrt{g} \,\tr\langle x|\log(\hat K) |x \rangle
\,.
\end{equation}
This brings in a pseudodifferential operator, namely, $\log(\hat K)$,
and its kernel at coincident points. Unfortunately, the logarithm does
not define an ultraviolet convergent (o even one-valued) operator for
any physical space-time dimension, correspondingly the kernel of
$\log(\hat K)$ diverges at coincident points, as also does its
trace. If the $\zeta$-function regularization is used
\cite{Dowker:1975tf,Hawking:1977ja,Elizalde:1994bk,Salcedo:1996qy},
this introduces a new pseudodifferential operator, the complex power
of the fluctuation operator, $({\hat K})^s$ \cite{Seeley:1967ea}. Its
kernel $\langle x|({\hat K})^s|y\rangle$ is an analytic entire
function with respect to $s$ provided the points $x$ and $y$ are
different. The diagonal matrix elements $\langle x|({\hat
K})^s|x\rangle$ are meromorphic functions of $s$ with a finite number
of poles which depend on the order of $\hat K$ and the space-time
dimension, but they are analytic at $s=0$. The computation of other
observables introduces further pseudodifferential operators ${\hat f}=
f(\nabla,M)$ and their diagonal matrix elements. For instance, for a
gauge current
\begin{equation}
\delta\Gamma= \int d^dx\sqrt{g} \,\tr(J^\mu(x)\delta A_\mu(x))
\end{equation}
with fluctuation operator of the Klein-Gordon type, ${\hat K} = 
-\nabla_\mu\nabla^\mu+M$, at one loop one formally obtains
\begin{equation}
J_{\text{1-loop}}^\mu(x)=-c\,\langle x|\left\{\nabla^\mu,(\hat K)^{-1}
\right\}|x\rangle\,,
\end{equation}
and again some regularization procedure has to be used to render the
expression meaningful.

The main purpose of this work is of practical and methodological
character, namely, to address the computation of diagonal matrix
elements of operators of the type ${\hat f}= f(\nabla,M)$. As we have
just shown such problem is ubiquitous in one-loop calculations in
quantum field theory. A more concrete goal is to extend methods
existing for flat space-time to curved space-time, the covariant
derivative carrying gauge and coordinate connections.

A useful technique when working with pseudodifferential operators is
the method of symbols
\cite{Seeley:1967ea,Eguchi:1980jx,Nepomechie:1984wt,Salcedo:1996qy}.
For an operator $\hat f$ constructed with $x^\mu$ and $\partial_\mu$
the symbol is essentially the function $f(x,p)$ such that ${\hat
f}=\,:\!f(x,\partial)\!:$, where the normal order stands for writing
$\partial_\mu$ at the right of $x^\mu$. Obviously the symbol is
closely related to the Wigner representation of operators
\cite{Wigner:1932eb} which is the basis of the phase space approach to
quantum mechanics \cite{Carruthers:1983fa}, except that Weyl normal
order \cite{Weyl:1927vd} is used instead, so that
$:\!f(x,\partial)\!:$ is Hermitian. As will be shown below, the symbol
allows to carry out manipulations, typically expansions of various
types, and directly or indirectly it has been used extensively in the
computation of the one loop effective action and related quantities,
such as the heat kernel
\cite{Schwinger:1951nm,Dewitt:1975ys,Gilkey:1975iq,%
Avramidi:2001ns,Kirsten:2002bk,Vassilevich:2003xt}, both in flat space-time
\cite{Nepomechie:1984wt,Chan:1986jq,Ball:1989xg,Salcedo:2004yh} and in
curved space-time
\cite{Barvinsky:1987uw,Gusynin:1989ky,Gusynin:1989nf,Avramidi:1991je,%
Avramidi:1994zp,Bel'kov:1996tn,vandeVen:1998pf,Moss:1999wq,%
Yajima:1988pj,Yajima:1995jk,Novozhilov:1991nm,Ceresole:1988hn}.  The
mathematical aspects of the symbol in Riemannian manifolds have been
considered e.g. in \cite{Pflaum:1996rp}. The extension of the method
of symbols to finite temperature field theory (in the imaginary time
formalism where space is compactified to a circle) has been carried
out in
\cite{Salcedo:1998sv,Garcia-Recio:2000gt,Megias:2002vr,Megias:2003ui}.
A further branch of mathematical physics where nowadays the symbol and
the Moyal product \cite{Moyal:1949sk} (which provides the symbol of
the product of two operators) have proven useful is in noncommutative
quantum field theory
\cite{Witten:1985cc,Konechny:2000dp,Douglas:2001ba}, where they play
an important role in the construction of noncommutative versions of
existing theories.

In general, the symbol is not a covariant quantity (either in the
gauge or coordinate senses) since the prescription $\partial_\mu\to
p_\mu$ is not covariant. Obviously, a naive covariant prescription of
the type $\nabla_\mu\to p_\mu$ would not define a faithful
representation of the pseudodifferential operator because, unlike
$p_\mu$, $\nabla_\mu$ is a non Abelian quantity. When the symbol is
used to compute a covariant quantity, such as the effective action,
covariance is recovered at the end of the computation, but is not
fully manifest in intermediate steps. Actually the situation is not as
bad as it seems (examples are given below) and it is usually not
necessary to go to the point of splitting $\nabla_\mu$ into
$\partial_\mu$ plus connections (thus spoiling the geometrical meaning
of the covariant derivative, a quite bold step to take even in the
simplest computations, particularly in the coordinate sector), but
nevertheless, it is a more or less severe nuissance. A possible way
out is to use ``covariant gauges'', namely, Fock-Schwinger
\cite{Fock:1937dy,Schwinger:1951nm} in the gauge sector and Riemann
normal coordinates
\cite{Alvarez-Gaume:1981hn,Muller:1997zk,Hatzinikitas:2000xe} in the
coordinate sector \cite{vandeVen:1998pf}. In \cite{Pletnev:1998yu}
Pletnev and Banin proposed a new method for the gauge sector in flat
space-time which implements the previous gauge fixing in a convenient
way. In the present work we name their construction {\em covariant
symbol} of the pseudodifferential operator. Whereas the ordinary
symbol is a function of $x^\mu$ and $p_\mu$, the covariant symbol is
actually an operator, but multiplicative in $x$-space, and hence
equivalent to a function of $x^\mu$. Unlike the ordinary symbol, however,
the covariant symbol is non multiplicative with respect to $p_\mu$,
that is, it contains $\partial/\partial p_\mu$. Eliminating
$\nabla_\mu$ at the price of introducing $\partial/\partial p_\mu$ is
in principle a net gain, though, since one is trading a non Abelian
quantity for an Abelian one. Besides being manifestly gauge covariant,
the covariant symbol has the interesting property that it is a
representation (in the technical sense of algebra homomorphism) of the
original operator. E.g., if $\overline f$ is the covariant symbol
$\hat f$, $\log(\overline f)$ is the covariant symbol of $\log({\hat
f})$. No Moyal product is required to compute the covariant symbol of
the product of two operators. Unfortunately there are also drawbacks:
in general, the covariant symbol cannot be computed in closed form
even for differential operators and thus expansions of some type are
usually required. In practice, this is not a serious disadvantage
since even the ordinary symbol of non differential operators is not
obtainable in closed form. In addition to manifest covariance, the
great virtue of the covariant symbol is that, due to their
homomorphism property, one needs to work them out for the building
blocks only, that is, the external fields $M$ and the covariant
derivative $\nabla_\mu$, and this can be done once and for all.  At
present, the method of covariant symbols has been successfully used in
several problems \cite{Pletnev:1998yu,Banin:2000qq,Banin:2001sv,%
Salcedo:2000hx,Salcedo:2002pr}.

In this work we take a step forward and extend the method of covariant
symbols to the case of curved space-time.\footnote{In their original
work \cite{Pletnev:1998yu} Pletnev and Banin proposed a formula
including the Riemannian connection, which, however, has not yet been
used in any actual application known to us. Our own proposal is
unrelated to that one.} The main issue now is to retain manifest
coordinate covariance, in addition to gauge covariance. In fact, we
find that working with the full covariant derivative as a whole (i.e.,
all connections included), as advocated for instance in
\cite{Vassilevich:2003xt}, is the cleanest way to proceed both
conceptually and computationally. To some extent the construction
carried out in the flat space-time case can be adapted to the curved
case. However, as is known, there are important technical differences
between gauge and coordinate cases. All differences stem from the fact
that the covariant derivative always adds a new coordinate index, and
thus a quantity $X$ and its covariant derivative $X_\mu=\nabla_\mu X$
fall in different representations of the group of diffeomorphisms.
This implies, for instance, that $\nabla_\nu$ acting on $X_\mu$ will
contain a further term $\Gamma_{\nu\mu}^\lambda$ not present in its
action on $X$. This is not so in the purely gauge case. When these
facts are properly taken into account, and with the help of Riemann
normal coordinates in an intermediate step, the construction in
\cite{Pletnev:1998yu} can be extended to the general case (i.e.,
coordinate plus gauge symmetries). The construction holds for
completely general connections in the world
sector\footnote{\label{foot:2} In this work we will use the label {\em
world} interchangeably with {\em coordinate} or {\em space-time} in
expressions like ``world tensor'', ``world index'', etc, to refer to
properties tied to indices $\mu,\nu,\ldots$, associated to natural
bases, $\partial/\partial x^\mu$, of the tangent space of the
space-time manifold.} (including e.g. torsion).

As we said the covariant symbol can seldom be obtained in closed
form. A natural expansion in this context is that in the number of
covariant derivatives (also known as adiabatic expansion), which
permits a systematic evaluation of the covariant symbol. For an
operator $\hat f= f(\nabla,M)$ it is sufficient to compute the
covariant symbols of $\nabla_\mu$ and $M$. We do this explicitly to
second order for a generic connection, and to fourth order for the
particular case of the Riemannian connection in the world sector (and
arbitrary connection in the gauge sector). The Laplacian is computed
to the same order in the derivative expansion. The computation
provides the covariant symbols in terms of elementary operators
classified by their number of derivatives. Such operators are just all
possible local covariant operators constructed with $M$, the field
strength tensor and their covariant derivatives. There is a number of
similarities between the covariant derivative expansion and the
standard heat kernel expansion. Both are expected to be asymptotic at
best. They are local and blind to global properties of the space-time
manifold. In both cases all possible local covariant operators are
expected to appear with some universal coefficients which are rational
numbers (to be determined by the computation). In the standard heat
kernel expansion the operators are classified by their (mass)
dimension. Because $\nabla_\mu$ has dimension 1, the heat kernel
expansion can be obtained a posteriori by means of a subsequent
reexpansion of the covariant derivative expansion
\cite{Salcedo:2004yh}.

The paper is organized as follows. In Section \ref{sec:2} we consider
flat space-time and revise the construction of ordinary and covariant
symbols in that case, as well as their use for the computation of
diagonal matrix elements. In Section \ref{sec:3} we discuss the
construction of ordinary symbols in curved space-time, highlighting
the subtleties introduced by the presence of curvature. In Section
\ref{sec:4} we extend covariant symbols to curved space-time, discuss
their properties and compute them to second order in the derivative
expansion for a general gauge and world connection. Section
\ref{sec:5} is devoted to set up a systematic computation of the
covariant symbols, and they are computed to fourth order for
Riemannian connection. In Section \ref{sec:6} we illustrate the ideas
and techniques involved by using the covariant symbols to explicitly
compute the diagonal matrix elements of a concrete operator to second
order. Finally, in Section \ref{sec:7} we present our conclusions. In
Appendix \ref{app:A} we summarize some of the conventions used in the
work. In Appendix \ref{app:B} we work out the same calculation as in
Section \ref{sec:6} but using the method of (ordinary) symbols. In
Appendix \ref{app:C} it is shown how to reduce momentum integrations
in curved space-time to those of the flat case.

\section{Symbols and covariant symbols in flat space-time}
\label{sec:2}

In this section we consider a $d$-dimensional flat space-time. The
operators act on states $\psi(x)$ which can be thought of as ``matter
fields'', as opposed to background external fields appearing in the
operators. In addition to their space-time dependence, the matter
fields may carry internal indices (however, for simplicity, we
disregard possible world indices in the fields throughout this
section). For concreteness, in what follows, we will assume that the
states are vectors in the fundamental representation of some internal
symmetry gauge group, and that the operators map them into the same
gauge representation. The scalar product takes the form $\langle
\psi_1|\psi_2\rangle= \int d^dx\,\psi_1^\dagger(x)\psi_2(x)$.

The pseudodifferential operators to be considered are of the form
$\hat f=f(D,M)$. They are constructed algebraically\footnote{This
means that $\hat f$ has the same algebraic properties as a sum of
products of $M$'s and $D$'s weighted with c-number coefficients, e.g.,
$D_\mu\log(D^2+M)$.}  out of the covariant derivative $D_\mu$ and one
or more multiplicative operators which are collectively denoted by
$M$. Such $M$ are just equivalent to matrix-valued (in internal
space) functions of $x$ acting as $M(x)\psi(x)$. The covariant
derivative is of the form $D_\mu=\partial_\mu+A_\mu(x)$, the gauge
connection $A_\mu(x)$ being also a matrix-valued function.

Under a gauge transformation $\psi(x)\to \Omega_g^{-1}(x)\psi(x)$, where
$\Omega_g$ is a multiplicative operator and $\Omega_g(x)$ a matrix in the
internal space. Correspondingly $M$, $D_\mu$ and $\hat f$ transform
under a similarity transformation
\begin{eqnarray}
&& M\to \Omega_g^{-1}M\Omega_g,
\nonumber\\
&& D_\mu\to \Omega_g^{-1}D_\mu\Omega_g,
\\
&& A_\mu\to \Omega_g^{-1}[\partial_\mu,\Omega_g]+
 \Omega_g^{-1}A_\mu\Omega_g,
\nonumber \\
&& \hat f\to \Omega_g^{-1}\hat f \, \Omega_g \,.
\nonumber
\end{eqnarray}
(The last equality being a consequence of the fact that $\hat f$ is
algebraically a function of $M$ and $D_\mu$.)

We can consider a basis of states of the form $|x,a\rangle$ with
spatial part equal to a Dirac delta located at $x$ and $a$ being a
gauge index, and the corresponding dual basis, $\langle
x,a|y,b\rangle=\delta(x-y)\delta^a_b$ . In what follows we will refer
to diagonal matrix elements of an operator $\hat f$ to mean those
matrix elements of the type $\langle x,a|\hat f|x,b\rangle$ ($a$ and
$b$ not necessarily equal). For convenience, we will occasionally
write the same matrix element omitting the internal indices,
i.e. $\langle x|\hat f|x\rangle$, and using a matrix notation in
internal space. The diagonal matrix element is gauge covariant,
\begin{eqnarray}
&& \langle x| \hat f|x\rangle \to \Omega_g^{-1}(x)\langle x| \hat f|x\rangle
\Omega_g(x),
\label{eq:2.2}
\end{eqnarray}
due to $\Omega_g|x\rangle=\Omega_g(x)|x\rangle$. Of course, this is
somewhat formal as $\langle x| \hat f|x\rangle$ does not exist for
many otherwise decent operators due to ultraviolet divergences in
taking the diagonal limit. Throughout this work we will assume that
the function $f$ is sufficiently convergent so that the matrix element
exists, or that a gauge invariant prescription, such as dimensional
renormalization or $\zeta$-function regularization has been used. Such
a prescription always exists for symmetries as gauge invariance, which
correspond to similarity transformations of $\hat f$.

\subsection{Symbol of an operator}
\label{sec:II.1}

In order to compute $\langle x| \hat f|x\rangle$, a standard technique
is the method of symbols \cite{Salcedo:1996qy}. Let $|0\rangle$ denote the 
wavefunction equal to one for all $x$, that is
\begin{eqnarray}
\langle x,a|0,b\rangle=\delta^a_b \,, \qquad
|0,a\rangle=\int d^dx|x,a\rangle .
\label{eq:2.3a}
\end{eqnarray}
Then, for a given point $x_0$,
\begin{eqnarray}
\langle x_0| f(D,M)|x_0\rangle
&=& \int d^dy\,\delta(y-x_0)\langle x_0|f(D,M)|y\rangle \nonumber\\
&=&
\int d^dy\frac{d^dp}{(2\pi)^d}e^{p(y-x_0)}\langle x_0|  f(D,M)|y\rangle
 \nonumber \\
&=&
\int d^dy\frac{d^dp}{(2\pi)^d}
\langle x_0|e^{-px}f(D,M)e^{px}|y\rangle
 \nonumber\\
&=&
\int \frac{d^dp}{(2\pi)^d}\langle x_0|e^{-px}f(D,M)e^{px}|0\rangle
 \nonumber\\
&=&
\int \frac{d^dp}{(2\pi)^d}\langle x_0|f(D+p,M)|0\rangle \,.
\label{eq:2.3}
\end{eqnarray}
In the second line $py:=p_\mu y^\mu$. Throughout this
work\footnote{Our notational conventions are summarized in Appendix
\ref{app:A}.} we will use an imaginary momentum variable $p_\mu=i{\rm
p}_\mu$ (${\rm p}_\mu$ real) to save unnecessary $i$ factors, however
$d^dp:=d^d{\rm p}$ is the standard integration in ${\mathbb{R}}^d$. In
the third line $x^\mu$ is the position operator. In the fourth line we
use the identity (\ref{eq:2.3a}). Finally, in the last line we use the
properties
\begin{equation}
e^{-px}M(x)e^{px}=M(x), \quad
e^{-px}\partial_\mu e^{px}= \partial_\mu+p_\mu \,.
\end{equation}

The quantity $\langle x|f(D+p,M)|0\rangle$ is known as the {\em
symbol} of the pseudodifferential operator $\hat f$. It is a
matrix-valued function of $x$ and $p$. For any multiplicative
operator, the property
\begin{equation}
\langle x_0|h(x)|0\rangle=h(x_0)
\label{eq:2.6a}
\end{equation}
implies that the symbol of the operator $M$ is just the function
$M(x)$. More generally, because $\partial_\mu|0\rangle=0$, the symbol
of $f(D,M)$ can be obtained by dragging the $\partial_\mu$'s to the
right and replacing them by $p_\mu$.

The matrix element $\langle x| f(D,M)|x\rangle$ is potentially
ultraviolet divergent. Using (\ref{eq:2.3}) the divergence is now
controlled by the momentum integration and one can make expansions or
other manipulations using the symbol of the operator. On the other
hand, $\langle x|f(D,M)|x\rangle$ is manifestly gauge covariant
(cf. (\ref{eq:2.2})) but the symbol is not, in general, due to the non
covariance of $|0\rangle$.  This implies that explicit gauge
covariance in (\ref{eq:2.3}) is only recovered after momentum
integration, but not in intermediate steps. Since this is an important
point let us dwell a bit on it. Clearly, if an operator $\hat h$ is
both covariant and multiplicative, the matrix element $\langle x|{\hat
h}|0\rangle$ will also be covariant\footnote{Indeed, $\langle x|{\hat
h}|0\rangle \to \langle x| \Omega_g^{-1}{\hat h}\Omega_g|0\rangle =
\Omega_g^{-1}(x){\hat h}(x) \Omega_g(x) = \Omega_g^{-1}(x)\langle
x|{\hat h}|0\rangle \Omega_g(x) $, since $\Omega_g^{-1}{\hat
h}\Omega_g$ is also a multiplicative operator.}. For instance, $\langle
x|D_\mu|0\rangle=A_\mu(x)$ (not covariant) whereas the matrix element
$\langle x|~|0\rangle$ of the multiplicative operator
$F_{\mu\nu}:=[D_\mu,D_\mu]$ is just $F_{\mu\nu}(x)=\partial_\mu
A_\nu-\partial_\nu A_\mu+[A_\mu, A_\nu]$ (covariant). So the lack of
covariance of the symbol stems from the fact that the operator
$f(D+p,M)$ (although covariant) is not multiplicative. On the other
hand, let
\begin{equation}
\hat{f}^\prime := \int \frac{d^dp}{(2\pi)^d}f(D+p,M) \,,
\label{eq:2.7}
\end{equation}
so that (\ref{eq:2.3}) becomes
\begin{equation}
\langle x_0| \hat{f}|x_0\rangle = \langle x_0|\hat{f}^\prime|0\rangle \,.
\end{equation}
The operator $\hat{f}^\prime$ is multiplicative (in addition to
covariant). This can be seen as follows.  For any (imaginary) constant
c-number $a_\mu$
\begin{equation}
e^{-ax}\hat{f}^\prime e^{ax}= \int \frac{d^dp}{(2\pi)^d}f(D+p+a,M) = \hat{f}^\prime \,,
\label{eq:2.9a}
\end{equation}
and this implies that $\hat{f}^\prime$ is multiplicative. The
operation in (\ref{eq:2.7}) projects the multiplicative component of
$\hat{f}$. Another observation is that, as can be seen from
(\ref{eq:2.9a}), an operator is multiplicative if an only if it is
invariant under the replacement $D_\mu\to D_\mu+a_\mu$, where $a_\mu$
is constant c-number, and in turn, this is true if and only if all
$D_\mu$ appear only in the form $[D_\mu,~]$. Some of these arguments
need to be modify in the curved case (see e.g. discussion relative to
$Z^0_{\mu\nu}$ in (\ref{eq:3.5a})).

A standard technique for using the symbol in the computation of
diagonal matrix elements of concrete operators (e.g. the heat kernel)
is as follows: $f(D+p,M)$ is expanded in powers of $D_\mu$ and
$M$. Each term so obtained is worked out by dragging the $D_\mu$ to
the right (or to the left) producing commutators of the type
$[D_\mu,~]$, which are gauge covariant and multiplicative. At the end,
there will be two type of summands, namely, i) those where all
$D_\mu$'s are inside commutators. These are multiplicative and so give
gauge covariant contributions to the symbol. And ii) summands where
the $D_\mu$ at the right cannot be arranged in commutators. These are
non multiplicative and break gauge covariance of the symbol. From the
previous discussion it follows that such terms cancel after momentum
integration and the surviving terms yield a covariant diagonal matrix
element.

The method just described is illustrated in Appendix \ref{app:B} for
the more involved case of curved space-time.\footnote{To revert the
calculation in Appendix \ref{app:B} to the flat space-time case
amounts to replace $\nabla_\mu$ with $D_\mu$, $g_{\mu\nu}$ with
$\delta_{\mu\nu}$, and to set to zero all $p_{\mu\nu\cdots}$ (having
two or more indices) as well as all Riemann tensors.\label{foot:1}}

\subsection{Covariant symbol}
\label{sec:II.2}

To achieve manifest gauge invariance prior to momentum integration in
(\ref{eq:2.3}), one can choose to work in the covariant Fock-Schwinger
gauge referred to the point $x_0$. An equivalent but more convenient
procedure was devised by Pletnev and Banin \cite{Pletnev:1998yu} who
introduced what we will call the (gauge) {\em covariant symbol} of an
operator. This is defined as follows
\begin{equation}
\overline{f}=e^{-\partial_p D}e^{-px}\hat f e^{px} e^{\partial_p D}
\label{eq:2.6}
\end{equation}
where $\partial_p D=\partial_p^\mu D_\mu$ and
$\partial_p^\mu=\partial/\partial p_\mu$. (Note that $\partial_p D=
D\partial_p $ just means the product of two operators, no derivative
of one on the other is implied.) Therefore, while the original
operator $\hat f$ acts on functions $\psi(x)$, its covariant symbol
$\overline{f}$ is an operator on functions $\psi(x,p)$. Key properties of
the covariant symbol are i) it is a multiplicative operator in $x$
space, ii) it is gauge covariant and iii) it is related to the
original operator by a similarity transformation.

Property iii) is obvious from its definition. ii) is also clear, since
$\hat f$, $x^\mu$, $p_\mu$, $\partial_p^\mu$, and $D_\mu$ are all
gauge covariant. Property i) holds provided the original operator
$\hat f$ does not contain $p_\mu$ and means that in the covariant
symbol all $\partial_\mu^x$ appear in commutators only. The
multiplicative property is equivalent to the statement
$[x^\mu,\overline{f}]=0$. That this is the case can be verified
directly from the definition. Alternatively, using the property
$\overline{\partial}_p^\mu=x^\mu$, which is easily verified, one has
\begin{equation}
[x^\mu,\overline{f}]= [\overline{\partial}_p^\mu,\overline{f}]
=\overline{[\partial_p^\mu,\hat{f}]}=0 \,.
\end{equation}
In the second equality we have used (iii). The multiplicative
property of the covariant symbol plays an important role in what
follows.

The covariant symbol can be used to compute the diagonal matrix element:
\begin{eqnarray}
\langle x_0| \hat{f}|x_0\rangle
&=&
\int \frac{d^dp}{(2\pi)^d}\langle x_0|e^{-px}\hat{f}e^{px}|0\rangle
 \nonumber\\
&=&
\int \frac{d^dp}{(2\pi)^d}\langle x_0|
e^{-\partial_p D}e^{-px}\hat f e^{px} e^{\partial_p D}|0\rangle 
 \nonumber\\
&=&
\int \frac{d^dp}{(2\pi)^d} \overline{f}(x_0) 
\,.
\label{eq:2.8}
\end{eqnarray}
The second equality follows from noting that $\partial_p^\mu$ at the
right vanishes (since it is acting on a $p$-independent wavefunction
in $(x,p)$ space). Likewise, $\partial_p^\mu$ at the left vanishes due
to integration by parts.\footnote{A more precise form of
(\ref{eq:2.8}) would be
$$
\langle x| \hat{f}|x\rangle=
\int \frac{d^dp}{(2\pi)^d}\langle x|\langle p| \overline{f}|0\rangle|0_p\rangle
= \langle x|\langle 0_p| \overline{f}|0\rangle|0_p\rangle \,,
$$ where $|0_p\rangle$ stands for
the unit wavefunction in $p$ space. The statement is then
$e^{\partial_p D}|0_p\rangle=|0_p\rangle$ and
$\langle0_p|e^{-\partial_p D}=\langle 0_p|$.\label{foot:4}} 
In the last equality we
have used that the covariant symbol is multiplicative to write
$\overline{f}(x_0)$ instead of the matrix element $\langle
x_0|\overline{f}|0\rangle$.

As we can see from the properties i-iii), the mapping
$\hat{f}\to\overline{f}$ really defines a representation of the
operators $\hat{f}$ in $x$-space in terms of multiplicative operators
(with respect to $x$) in $(x,p)$-space which is consistent with gauge
covariance. A further property is that the covariant symbol preserves
the hermiticity properties of the original operator.

Because the covariant symbol is a representation (i.e. an algebra
homomorphism) one has e.g.
\begin{eqnarray}
\overline{f(D,M)}=f(\overline{D},\overline{M}) \,,
\nonumber\\
\overline{F}_{\mu\nu}=[\overline{D}_\mu,\overline{D}_\nu] \,,
\nonumber\\
\overline{[D_\mu,M]}=[\overline{D}_\mu,\overline{M}] \,,
\label{eq:2.9}
\end{eqnarray}
and so on. Using the definition (\ref{eq:2.6}), the basic operators
$\overline{D}_\mu$ and $\overline{M}$ can be readily computed in terms of a
covariant derivative expansion \cite{Pletnev:1998yu}
\begin{eqnarray}
{\overline{M}} &=& 
e^{-[\partial_p D,~]}M
\nonumber \\
&=&
M -M_\mu\,\partial_p^\mu
+\frac{1}{2!}M_{\nu\mu}\,\partial_p^\nu\partial_p^\mu
-\frac{1}{3!}M_{\alpha\nu\mu}\,\partial_p^\alpha\partial_p^\nu\partial_p^\mu
+\cdots \,, \nonumber \\ 
{\overline{D}_\lambda} &=& 
e^{-[\partial_p D,~]}(D_\lambda+p_\lambda)
\nonumber
 \\
&=&
 p_\lambda
-\frac{1}{2!}F_{\mu\lambda}\,\partial_p^\mu
+\frac{2}{3!}F_{\nu\mu\lambda}\,\partial_p^\nu\partial_p^\mu
-\frac{3}{4!}F_{\alpha\nu\mu\lambda}\,\partial_p^\alpha\partial_p^\nu
\partial_p^\mu
+\cdots \,.
\label{eq:2.10}
\end{eqnarray}
In writing this formulas we have denoted the derivatives of $M$ and
$F_{\mu\nu}$ by introducing the convention $[D_\mu,X_I]=X_{\mu I}$,
i.e.,
\begin{equation}
[D_\mu,X_{\alpha_1\cdots\alpha_n}]=X_{\mu\alpha_1\cdots\alpha_n} \,.
\end{equation}
As is readily verified, the expansions in (\ref{eq:2.10}) are
consistent with the last two equations in (\ref{eq:2.9}). (${\overline
F}_{\mu\nu}$ follows the same formula as ${\overline M }$ since the
latter only assume $M$ to be a multiplicative operator.)

As a further convention, we will exploit the fact that the derivatives
$\partial_p^\mu$ commute and so all their indices are symmetrized, to
use a single symbol $s$ for all of them, that is, we will often write
\begin{eqnarray}
{\overline{M}} &=& 
M -M_s\,\partial_p^s
+\frac{1}{2!}M_{ss}\,(\partial_p^s)^2
-\frac{1}{3!}M_{sss}\,(\partial_p^s)^3
+\cdots \,, \nonumber \\ 
{\overline{D}_\lambda} &=& 
 p_\lambda
-\frac{1}{2!}F_{s\lambda}\,\partial_p^s
+\frac{2}{3!}F_{ss\lambda}\,(\partial_p^s)^2
-\frac{3}{4!}F_{sss\lambda}\,(\partial_p^s)^3
+\cdots \,.
\label{eq:2.12}
\end{eqnarray}

The use of covariant symbols to compute diagonal matrix elements is
illustrated in Section \ref{sec:6} for curved space-time. Since it is
easy to reduce that calculation to the simpler case of flat space-time
(see footnote \ref{foot:1}) we do give further examples here.

Note that the covariant symbol method is compatible with derivative
expansions (see \cite{Salcedo:2000hx,Salcedo:2000hp} for strict
derivative expansions of the effective action functional of Dirac
fermions using this method). Such expansions are expected to be
asymptotic in general.

Another comment has to do with momentum space integration by
parts. Formally, $f(D+p,M)$ and ${\overline{f}}= e^{-\partial_p
D}f(D+p,M)e^{\partial_p D}$ would differ by terms with $\partial_p$,
implying that the difference should vanish on $|0_p\rangle$ or
$\langle 0_p|$ (see footnote \ref{foot:4}). This formal argument is correct
for sufficiently well behaved operators in the ultraviolet, e.g. $M$
but not for $D_\mu$ in (\ref{eq:2.12}), (see also
$\overline{\nabla_\mu\nabla^\mu}{}^{(2)}$ in (\ref{eq:5.38})). Of
course, it is never necessary to take diagonal matrix elements of
divergent operators (without some regularization to make them
convergent). The homomorphism property implies that
${\overline{D}_\lambda}$ provides the suitable momentum dependence to
give the correct result when used as part of an ultraviolet convergent
operator.

To summarize this section, the ordinary symbols are representations of
pseudodifferential operators in terms of functions of $x$ and $p$
which are matrix valued in internal space and they are not gauge
covariant. The representation introduced by Pletnev and Banin in flat
space-time, on the other hand, is in terms of operators which are
multiplicative with respect to $x$ and so equivalent to functions. In
this sense they are similar to the ordinary symbols (which also remain
operators in internal space). They are covariant and enjoy the
homomorphism property, at the price of being non multiplicative in $p$
space.  Both ordinary and covariant symbols provide diagonal matrix
elements upon integration over $p$.

\section{Symbols in curved space-time}
\label{sec:3}

\subsection{General considerations}

The method of symbols can be extended to curved space-time. The main
issue now is to preserve both gauge covariance and coordinate or
world\footnote{See footnote \ref{foot:2}.}  covariance. The
space-times we consider may have Euclidean or Minkowskian
signatures. We will treat the two cases simultaneously since there is
no formal difference for our purposes. We will often refer to the
Riemannian connection to mean the unique torsionless metric preserving
connection, regardless of the signature of the metric.

The pseudodifferential operator is now of the form
$\hat{f}=f(\nabla,M)$ where $\nabla_\mu$ is the covariant derivative
and includes connections for the parallel transport of all indices:
world and internal indices. The latter include gauge, Lorentz frame,
Dirac indices in the case of fermions, and so on.\footnote{We are
following the approach found for instance in
\cite{Vassilevich:2003xt}. In this approach, if $e_\mu^a$ is the
tetrad field, the connections on the indices $\mu$ and $a$ are such
that $\nabla_\nu e_\mu^a=0$, likewise for the Dirac gammas,
$\nabla_\mu\gamma^\nu=0$ with a suitable connection acting on the
Dirac indices. This convention is not universally adopted. For
instance, in \cite{Alvarez-Gaume:1985dr} $\nabla_\nu e_\mu^a$ would
only include the connection for the world index while $D_\nu e_\mu^a$
would only include the connection for the tetrad index.} In what
follows we use indifferently ``internal'' or ``gauge'' index to mean
any kind of internal index. The matter fields $\psi(x)$ may contain
internal indices as well as world indices. Likewise, the external
fields $M$ may also contain all kind of indices and act as
multiplicative operators with respect to $x$. The metric
$g_{\mu\nu}(x)$ is an example of such a field.

We do not assume that $f(\nabla,M)$ should be a world scalar
(cf. Section \ref{sec:6} for an example). As a consequence $\hat{f}$
may connect different diffeomorphism representations. The reason for
this generality is that there is no net gain in restricting oneself to
the equal representation case. This is because we need to consider
covariant symbols not only of the final operator $\hat{f}$ but also of
$\nabla_\mu$ as a building block and $\nabla_\mu$ always connects
different tensor representations. This is an important difference with
the gauge case where one can work consistently viewing all operators
as matrices in internal space.\footnote{In the gauge case (and flat
space-time), if $\psi$ is a gauge vector $D_\mu\psi$ is again a gauge
vector and so in a second derivative $D_\nu D_\mu\psi$, $D_\nu$ would
still be ``the same operator'' $\partial_\nu+A_\nu$. In the general
covariant case (and gauge singlet), if $\phi$ is a world scalar,
$\nabla_\mu\phi=\partial_\mu\phi$ is a world vector and $\nabla_\nu$
``acts differently'' on it, namely, as $\nabla_\nu=
\partial_\nu-\Gamma_\nu$ ($\Gamma_\nu$ being a matrix on world indices
$(\Gamma_\nu)_\alpha{}^\beta= \Gamma_{\nu\alpha}^\beta$). The point
is, of course, that due to its geometrical meaning (through the
Leibnitz rule), $\nabla_\mu$ acts consistently at each place
(i.e. using the proper connection) and there is no need to worry about
such details. \label{foot:6}} Of course, it would be pointless to try
to erase the difference between gauge and coordinate cases using a
tetrad, e.g. $D_a=e^\mu_a\nabla_\mu$ \cite{Weinberg:1972bk} since,
although $D_a$ is a world scalar, it connects now different internal
representations (namely, with respect the new internal structure
introduced by the Lorentz index $a$).

For states in equal representations the scalar product
is
\begin{equation}
\langle \psi_1|\psi_2\rangle= \int d^dx\sqrt{g(x)}\,\psi_1^\dagger(x)\psi_2(x) \,,
\end{equation}
$g(x)$ denoting $|\det g_{\mu\nu}|$. As usual the scalar product has
been defined so that it is coordinate invariant, although metric
dependent. (Note however that, as shown below, the construction of the
covariant symbols themselves do not require a metric to be defined.)
For states in different representations the scalar product
vanishes. An active world (or coordinate, or diffeomorphism)
transformation, $x^\mu\to x^{\prime\mu}(x)$ defines a corresponding
operator on states $\psi\to \hat\Omega_w^{-1}\psi$, which takes the
form
\begin{equation}
\psi(x)\to (\hat\Omega_w^{-1}\psi)(x)=\psi(x^\prime(x))
\end{equation}
for a scalar,
\begin{equation}
\psi_\mu(x)\to \frac{\partial 
x^{\prime\alpha}}{\partial x^\mu}\psi_\alpha(x^\prime(x))
\end{equation}
on covariant world vectors, and so on.

As in the flat case we will use a basis of the tensor product type,
with states $|x_0,a,w\rangle$ located at $x_0$ (wavefunction
$\delta(x-x_0)/\sqrt{g(x)}$), $a$ being a gauge index and $w$ a set of
world indices (empty for world scalar states), with dual basis
$\langle x,a,w|y,b,w^\prime\rangle=
\delta^a_b\delta^w_{w^\prime}\delta(x-y)/\sqrt{g(x)}$. The metric in
the space-time factor of the basis states is introduced so that they
are world scalars, and similarly for the scalar product.

Once again we want to evaluate diagonal (in $x$) matrix elements
$\langle x,a,w|\hat{f}|x,b,w^\prime\rangle$. For short we will often
write just $\langle x|\hat{f}|x\rangle$, however, one should keep in
mind the presence of world and internal indices since they determine
how the covariant derivative acts. In particular, the operator
$\hat{f}$ should connect the in representation $(b,w^\prime)$ with the
out representation $(a,w)$, so that $\langle
\psi_1|\hat{f}|\psi_2\rangle$ is a gauge and world singlet.

As in the purely gauge case, we can regard the matrix element $\langle
x|f(\nabla,M)|x\rangle$ as a (gauge and world) covariant function of
$x$ which takes values on operators acting on internal and world
indices.  And in turn this can be viewed as equivalent to a
covariant multiplicative operator (in the purely gauge case, such multiplicative
operator is $\hat{f}^\prime$ introduced in (\ref{eq:2.7})).

Because multiplicative operators play an important role in what
follows let us define them more precisely. A {\em c-number
multiplicative operator}, ${\hat \phi}$, is one that acts in the form
\begin{equation}
{\hat \phi}|x,a,w\rangle = \phi(x)|x,a,w\rangle
\label{eq:3.2-1}
\end{equation}
$\phi(x)$ being a fixed complex function. Such operator is gauge singlet and
world scalar. Now, by definition a {\em multiplicative operator} is
one that commutes with all c-number multiplicative operators. A
multiplicative operator is diagonal in $x$ but in general non diagonal
with respect to all other degrees of freedom. In commutators, the
c-number multiplicative operators are blind to those degrees of
freedom but sensible to derivatives with respect to $x$. Here we can
see a difference between the purely gauge case and the general case
(gauge plus world degrees of freedom). In the flat case,
$[D_\mu,D_\nu]$ and $[D_\mu,{\cal Q}]$, with ${\cal Q}$
multiplicative, are multiplicative operators, as is easily
verified. This property is lost when a world connection is included,
i.e. for $\nabla_\mu$. For a generic world connection, the operator
\begin{equation}
Z^0_{\mu\nu}:= [\nabla_\mu,\nabla_\nu]
\label{eq:3.5a}
\end{equation}
is not multiplicative, since acting on a world scalar and gauge
singlet state $\phi(x)$, it gives
\footnote{Our conventions are such that, for a world vector gauge singlet
$V^\lambda$
$$
[\nabla_\mu,\nabla_\nu] V^\lambda=
+R_{\mu\nu~\sigma}^{~~\,\,\lambda~}V^\sigma
-T_{\mu\nu}^{~~\,\sigma}\nabla_\sigma V^\lambda \,,\quad
\Rc_{\mu\nu}:=R_{\lambda\mu~\nu}^{~~\,\,\lambda~} \,,\quad
\R:=\Rc^\lambda_{~\,\lambda} \,.
$$
}
\begin{equation}
[\nabla_\mu,\nabla_\nu] \phi= -T_{\mu\nu}^{~~\,\lambda}\nabla_\lambda
\phi 
= -T_{\mu\nu}^{~~\,\lambda}\frac{\partial\phi}{\partial x^\lambda}
\,.
\end{equation}
being $T_{\mu\nu}^{~~\,\lambda}$ the torsion. (Equivalently,
$[[\nabla_\mu,\nabla_\nu],\hat\phi]=
-T_{\mu\nu}^{~~\,\lambda}[\nabla_\lambda, \hat\phi]$, in terms of
c-number multiplicative operators.)  The result depends on derivatives
of $\phi$ and so such $Z^0_{\mu\nu}$ is not a multiplicative operator
in the presence of torsion. A remedy is to introduce the new operator
\begin{equation}
Z_{\mu\nu}:=[\nabla_\mu,\nabla_\nu]
+\frac{1}{2}\{\nabla_\lambda,T_{\mu\nu}^{~~\,\lambda}\}
\end{equation} 
($\{\,,\}$ denotes anticommutator) which is multiplicative, as is
readily verified. It coincides with $Z^0_{\mu\nu}$ for a torsionless
connection such as the Levi-Civita or Riemannian connection (in the
world sector). Nevertheless, $[\nabla_\alpha,Z_{\mu\nu}]$ is again non
multiplicative (even for the Riemannian connection). In addition, at
variance with $F_{\mu\nu}$ of the flat case, $Z_{\mu\nu}$ will not
commute with the momentum $p_\mu$ (to be introduced subsequently,
similar to the flat case). This is because $Z_{\mu\nu}$ acts on world
indices
\begin{equation}
[Z_{\mu\nu}, p_\lambda]= -R_{\mu\nu~\,\lambda}^{~~~\sigma}\,p_\sigma 
\,.
\end{equation}
(This formula holds for any world connection.) As noted previously, the
difference between $D_\mu$ and $\nabla_\mu$ is due to fact that
$\nabla_\mu$ acts on world indices but also adds world indices.

Because there are several types of quantities to be considered, we
will introduce the following notation: the more general quantities or
objects (such as operators, wavefunctions, matrix elements, etc) to be
considered in this Section belong to the class
\begin{equation}
{\cal C}(x,\nabla,Z,W,I,p)\,.
\end{equation}
The presence of the label ($x$) indicates that the quantity in
question may depend on $x^\mu$. Likewise, the label ($\nabla$) denotes
that the object may be non multiplicative in $x$ space. ($Z$) means
that it may contain $Z_{\mu\nu}$ or other {\em multiplicative
operators that act on world indices}. ($W$) indicates that it may
contain world indices. ($I$) that they may contain internal (or gauge,
or bundle) indices. Finally, ($p$) that it may depend on $p_\mu$.  On
the other hand, the class
\begin{equation}
{\cal C}(x,\underline{\nabla},Z,W,\underline{I},p)\,,
\end{equation}
or simply $ {\cal C}(\underline{\nabla},\underline{I})$, will indicate
quantities which are multiplicative in $x$ space [do not contain
``free'' $\nabla_\mu$'s] (denoted with $\underline{\nabla}$) and are
gauge singlets [do not contain internal indices] (denoted with
$\underline{I}$). And similarly for other underlined labels. Thus, for
instance, $p_\mu$ is in class 
${\cal C}(\underline{\nabla},\underline{Z},\underline{I})$,
the operators $M$ in $f(\nabla,M)$ are in class 
${\cal C}(\underline{\nabla},\underline{Z},\underline{p})$, and $Z_{\mu\nu}$
is in class $ {\cal C}(\underline{\nabla},\underline{p})$ (while
$Z^0_{\mu\nu}\in{\cal C}(\underline{p})$ for a world connection with
torsion). c-number multiplicative operators are in
${\cal C}(\underline{\nabla},\underline{Z},\underline{I},\underline{W})$.
Multiplicative operators are in ${\cal C}(\underline{\nabla})$.

\subsection{Diagonal matrix elements}

To implement the method of symbols as for the flat case, we proceed
similarly to (\ref{eq:2.3}), starting with the diagonal matrix element
\begin{eqnarray}
\langle x_0| \hat{f}|x_0\rangle
&=& \frac{1}{\sqrt{g(x_0)}}
\int d^dx_1\delta(x_1-x_0)\langle x_0|f(\nabla,M)|x_1\rangle\sqrt{g(x_1)}
 \nonumber\\
&=&
\frac{1}{\sqrt{g(x_0)}}
\int d^dx_1\frac{d^dp}{(2\pi)^d}e^{p(x_1-x_0)}
 \langle x_0|  f(\nabla,M)|x_1\rangle\sqrt{g(x_1)}
\label{eq:3.2a}
\end{eqnarray}
The matrix element is independent of the coordinate system, but the
symbol is not. So we will pick a certain {\em reference coordinate
system} (RCS) and denote its coordinates by $\xi^A(x)$,
$A=1,\ldots,d$. These are $d$ world scalar functions, furthermore, we
set to zero the connection associated to the indices $A$ (recall that
$\nabla_\mu$ is defined acting on all indices with the appropriate
connection). We will reserve the symbol $x^\mu$ to denote an arbitrary
coordinate system. Then
\begin{equation}
\nabla_\mu\xi^A= \frac{\partial\xi^A}{\partial x^\mu} =: t^A_\mu(x) \,.
\end{equation}
These are world vector. Let us also introduce the dual contravariant world vectors
\begin{equation}
t^\mu_A(x):=\frac{\partial x^\mu}{\partial\xi^A}
\end{equation}
such that
\begin{equation}
t^A_\mu \, t^\mu_B=\delta^A_B\,,\quad
t_A^\mu \, t_\nu^A=\delta^\mu_\nu\,.
\label{eq:5.2}
\end{equation}

Using in (\ref{eq:3.2a}) the RCS, if $\xi^A_0$, $\xi^A_1$, and
$g^{(\xi)}(x)$, denote, respectively, the coordinates of the point
$x_0$ and $x_1$, and the determinant of the metric in the RCS, we find
\begin{eqnarray}
\langle x_0|{\hat f}|x_0\rangle
&=&
\frac{1}{\sqrt{g^{(\xi)}(x_0)}}
\int d^d\xi_1\frac{d^dp_A}{(2\pi)^d}e^{p_A(\xi_1^A-\xi_0^A)}
 \langle x_0|  f(\nabla,M)|x_1\rangle\sqrt{g^{(\xi)}(x_1)} \,.
\end{eqnarray}
The momentum integration variables $p_A$ are $d$ c-number and
space-time constant quantities, $p_A\in{\cal
C}(\underline{x},\underline{\nabla},\underline{Z},\underline{W},\underline{I},p)$.

It is convenient to work in an arbitrary coordinate system. To this end, let us
define the world vector fields
\begin{equation}
p_\mu=t_\mu^Ap_A \,,
\qquad
X^\mu=t^\mu_A\xi^A \,.
\end{equation}
The c-number function
$\Phi=p_A\xi^A$ can also be written as $p_\mu X^\mu$, and moreover
\begin{equation}
p_\mu = \nabla_\mu\Phi \,.
\end{equation}
We can write
\begin{eqnarray}
\langle x_0| {\hat f}|x_0\rangle
&=&
\frac{1}{\sqrt{g^{(\xi)}(x_0)}}
\int d^dx_1\frac{d^dp_A}{(2\pi)^d}e^{p_\mu(X_1^\mu-X_0^\mu)}
\langle x_0|  f(\nabla,M)|x_1\rangle \sqrt{g(x_1)} \,.
\label{eq:3.2b}
\end{eqnarray}
Now, being $p_\mu X^\mu$ a c-number multiplicative operator, we can
apply (\ref{eq:3.2-1}) to write
\begin{eqnarray}
\langle x_0|\hat{f}|x_0\rangle
&=&
\frac{1}{\sqrt{g^{(\xi)}(x_0)}}
\int d^dx_1\frac{d^dp_A}{(2\pi)^d}
\langle x_0|e^{-pX} f(\nabla,M)e^{pX}|x_1\rangle \sqrt{g(x_1)} \,.
\label{eq:3.2c}
\end{eqnarray}
Introducing now the space-time constant states $|0,a,w\rangle$, which
lie in the class ${\cal C}(\underline{x},\underline{\nabla},
\underline{Z},\underline{p})$,

\begin{eqnarray}
\langle x,a,w|0,b,w^\prime\rangle=\delta^a_b\delta^w_{w^\prime} \,, \qquad
|0,a,w\rangle=\int d^dx\sqrt{g(x)}|x,a,w\rangle \,,
\end{eqnarray}
yields
\begin{eqnarray}
\langle x_0|\hat{f}|x_0\rangle
&=&
\frac{1}{\sqrt{g^{(\xi)}(x_0)}}
\int \frac{d^dp_A}{(2\pi)^d}
\langle x_0|e^{-pX}f(\nabla,M)e^{pX}|0\rangle
 \nonumber\\
&=&
\frac{1}{\sqrt{g^{(\xi)}(x_0)}}
\int \frac{d^dp_A}{(2\pi)^d}
\langle x_0|f(\nabla+p,M)|0\rangle \,,
\label{eq:3.2d}
\end{eqnarray}
where we have used the identity (understood as product of three operators)
\begin{equation}
e^{-p X}\nabla_\mu e^{p X}= \nabla_\mu +p_\mu \,.
\end{equation}

The quantity $\langle x_0|f(\nabla+p,M)|0\rangle$ is now the symbol of
$\hat{f}$ at $x_0$. Again it is not gauge covariant, since under a
local gauge transformation, $\Omega_g(x)$, $|0\rangle$ stops being
space-time constant. For a similar reason, it is not general covariant
under $\hat\Omega_w$, unless $|0\rangle$ is a world scalar. (However,
as in the gauge case, covariance is recovered for matrix elements
$\langle x_0|~|0\rangle$ of multiplicative operators.) In addition,
the symbol depends on the choice of RCS through $p_\mu$ and $t_\mu^A$.
$p_\mu$ is the vector field which happens to take constant components
$p_A$ in the RCS, and similarly $X^\mu$ has components $\xi^A$
precisely in that coordinate system. Such RCS dependence cancels after
momentum integration, as $\langle x_0|f(\nabla,M)|x_0\rangle$ is gauge
and coordinate covariant. As in the flat space-time case
(\ref{eq:3.2d}) can also be written as
\begin{equation}
\langle x_0| \hat{f}|x_0\rangle = \langle x_0|\hat{f}^\prime|0\rangle
\end{equation}
with
\begin{equation}
\hat{f}^\prime := \frac{1}{\sqrt{g^{(\xi)}(x)}}\int \frac{d^dp_A}{(2\pi)^d}f(\nabla+p,M) \,,
\label{eq:2.7a}
\end{equation}
($g^{(\xi)}(x)$ being a c-number multiplicative operator here.)
The operator $\hat{f}^\prime$ is multiplicative, formally independent
of the in and out state spaces, and gauge and world
covariant. Furthermore, it is RCS independent; the RCS dependence of
the momentum integral through the vector field $t^A_\mu$ in $p_\mu$
exactly cancels with the prefactor $1/\sqrt{g^{(\xi)}}$ (cf. Appendix
\ref{app:C}). As the scalar product itself, $\hat{f}^\prime$ is metric dependent.

As in the flat case, we can work out $f(\nabla+p,M)$ by dragging
$\nabla_\mu$ to the right.\footnote{And the rule
$\partial^x_\mu|0\rangle=0$ still applies. However, as in the purely
gauge case, it is not practical to lose manifest covariance. It is
preferable to let the momentum integration to kill non multiplicative
(and so non covariant) contributions. See Appendix \ref{app:B} for an example.} A
very important difference with the flat case is that $\nabla_\mu$ and
$p_\nu$ no longer commute. Their commutator is just the covariant
derivative of $p_\mu$,
\begin{equation}
[\nabla_\mu,p_\nu]=[\nabla_\mu,t_\nu^A]p_A=t_{\mu\nu}^Ap_A
=t^\lambda_A t_{\mu\nu}^A\,p_\lambda
\end{equation}
($t_{\mu\nu}^A$ being the covariant derivative of $t_\nu^A$, according
to our convention.)  The same computation in the RCS system, where
$p_\mu$ equals $p_A$ (and so $\partial^x_\mu p_\nu=0$), gives
\begin{equation}
[\nabla_\mu,p_\nu]= -\Gamma^{(\xi)}{}_{\mu\nu}^\lambda p_\lambda \,.
\end{equation}
In an arbitrary coordinate system this becomes
\begin{equation}
[\nabla_\mu,p_\nu]= -P_{\mu~\,\nu}^{~\lambda} p_\lambda \,,
\label{eq:3.5}
\end{equation}
where $P_{\mu~\,\nu}^{~\lambda}$ is the world tensor with components
precisely equal to $\Gamma_{\mu\nu}^\lambda$ in the RCS. Hence,
\begin{equation}
P_{\mu~\,\nu}^{~\lambda}= -t^\lambda_A t_{\mu\nu}^A \,.
\label{eq:3.6}
\end{equation}

As we mentioned before, another difference with the flat case (and hence
another complication) is that the construction $[\nabla_\mu,~]$ does
not automatically produce multiplicative operators. In any case, after
moving the $\nabla$'s to the right, in principle one will be able to
manage to form multiplicative combinations of nablas in some terms,
plus terms in which non multiplicative combinations appear at the
right. The latter vanish upon momentum integration. Such integration
will usually require to put all $p_\mu$'s {\em together} at the left
(recall that $Z_{\mu\nu}$ and $p_\lambda$ do not commute) except those
in the form $p_\mu p^\mu$, appearing in propagators, etc. Note that
$p_\mu p^\mu$ is not constant even in the RCS because
$p^\mu(x)=g^{\mu\nu}(x)p_\nu$, however, it is a c-number
multiplicative operator, so it commutes with all multiplicative
operators ($p_\mu p^\mu$ fails to commute with non multiplicative
operators, but those have been already been disposed of.) The key point
is that when only multiplicative operators appear, the tensor
$P_{\mu~\,\nu}^{~\lambda}$ and its covariant derivatives will be
needed only at the point $x=x_0$. Upon momentum integration these
tensors will appear only through combinations which are independent of
the choice of the RCS, e.g.
\begin{equation}
\nabla_\mu P_{\nu~\,\beta}^{~\alpha}-\nabla_\nu P_{\mu~\,\beta}^{~\alpha}
-P_{\mu~\,\lambda}^{~\alpha}P_{\nu~\,\beta}^{~\lambda}
+P_{\nu~\,\lambda}^{~\alpha}P_{\mu~\,\beta}^{~\lambda}
=R_{\mu\nu~\,\beta}^{~~\,\,\alpha} \,.
\end{equation}
(For the momentum integration with $x$-dependent $p_\mu p^\mu$ see
Appendix \ref{app:C}.) In practice, the natural way to proceed is to
take as RCS the Riemann normal coordinates at $x=x_0$ from the
beginning since this choice provides manifestly covariant results for
the tensor $P_{\mu~\,\nu}^{~\lambda}$ and its covariant
derivatives.

In Appendix \ref{app:B} we illustrate all previous points for the
operator ${\hat Q}_{\mu\nu}$ in (\ref{eq:6.1}) and the Riemann
connection. The matrix elements of this operator are computed to
second order in a derivative expansion using the (ordinary) symbols
method.

\section{Covariant symbols in curved space-time}
\label{sec:4}

As we have just sketched in the previous Section, one can work with
the ordinary symbols for pseudodifferential operators in curved
space-time along the same lines as for the flat case, although things
are, in general, more involved in the curve case and covariance is
recovered only after momentum integration. In this Section we
introduce the covariant symbols in presence of curvature. They are
fully covariant representations in terms of operators which are
multiplicative with respect to $x$.

We will need derivatives with respect to the momenta $p_A$. These are denoted as $\partial^A$
\begin{equation}
\partial^A:= \frac{\partial}{\partial p_A} \,,
\quad
\partial^\mu:= t_A^\mu \, \partial^A \,,\quad
\end{equation}
The derivatives $\partial^\mu$ are contravariant world vectors which satisfy
\begin{equation}
[\partial^\mu,p_\nu]:= \delta^\mu_\nu
\end{equation}
and, as a consequence of (\ref{eq:3.5})
\begin{equation}
[\nabla_\mu,\partial^\nu]= P_{\mu~\,\lambda}^{~\nu} \partial^\lambda \,.
\label{eq:4.3a}
\end{equation}

We should extend now our previous notation. The most general objects belong to the class
\begin{equation}
{\cal C}(x,\nabla,Z,W,I,p,\partial)\,,
\end{equation}
where the new label $\partial$ indicates a possible dependence on
$\partial^A$.\footnote{From now on we use $\partial^\mu$ to denote
$\partial^\mu_p$ since the non covariant operator
$\partial_\mu^x=\partial/\partial x^\mu$ will appear rarely.}  On the
other hand the class, $ {\cal C}({\underline\partial})$ denotes
quantities which are multiplicative with respect to $p_A$.

In the curved case we introduce a preliminary definition of the
covariant symbol of an operator $\hat{f}$ as
\begin{equation}
\overline{f}:=
e^{-\frac{1}{2}\{\nabla_\mu,\partial^\mu\}}
e^{-p_\alpha X^\alpha}
\hat{f}\,
e^{p_\beta X^\beta}
e^{\frac{1}{2}\{\nabla_\nu,\partial^\nu\}}
\,,\qquad
\hat{f}\in {\cal C}(\underline{p},
\underline{\partial}) \,.
\label{eq:4.3}
\end{equation}
This definition, as well as the general properties to be discussed
below, holds actually for any connection on the world indices,
although eventually we will restrict ourselves to the Riemannian connection for
simplicity. As shown subsequently, the map 
$\hat{f}\mapsto \overline{f}$ defines an operator representation from
${\cal C}(x,\nabla,Z,W,I,\underline{p},\underline{\partial})$ into
${\cal C}(x,\underline{\nabla},Z,W,I,p,\partial)$.

The definition depends on the choice of the (arbitrary) RCS (in which
the tensors $p_\mu$ and $\partial^\mu$ have constant
components). Eventually we will take the RCS as the Riemann normal
coordinate system, thereby obtaining fully covariant expressions for
the covariant symbol. Because $\nabla_\mu$ and $\partial^\mu$ do not
commute (cf. (\ref{eq:4.3a})), one can extend the construction
$e^{D_\mu\partial^\mu}$ corresponding to the flat case in several
different ways, among other, as $e^{\nabla_\mu\partial^\mu}$ or
$e^{\partial^\mu\nabla_\mu}$, or even as
$e^{\frac{1}{2}\{\nabla_\mu,\partial^\mu\}}$. All of them are
valid. The two former choices give slightly simpler formulas, but the
latter has the virtue of preserving the hermiticity properties of the
original operator, $e^{pX}$ being unitary.

The use of the covariant symbol to compute the diagonal matrix element is
fully analogous to its flat space-time version
\begin{eqnarray}
\langle x_0| \hat{f}|x_0\rangle
&=&
\frac{1}{\sqrt{g^{(\xi)}(x_0)}}
\int \frac{d^dp_A}{(2\pi)^d}\langle x_0|e^{-pX}\hat{f}e^{pX}|0\rangle
 \nonumber\\
&=&
\frac{1}{\sqrt{g^{(\xi)}(x_0)}}
\int \frac{d^dp_A}{(2\pi)^d}\langle x_0|
e^{-\frac{1}{2}\{\nabla,\partial\}}
e^{-pX}\hat{f}e^{pX}
e^{\frac{1}{2}\{\nabla,\partial\}}|0\rangle
 \nonumber\\
&=&
\frac{1}{\sqrt{g^{(\xi)}(x_0)}}
\int \frac{d^dp_A}{(2\pi)^d} \, \langle x_0|\overline{f}|0\rangle
\,.
\label{eq:4.5}
\end{eqnarray}
In the second equality we have used
\begin{equation}
\frac{1}{2} \{\nabla_\mu,\partial^\mu\}
 =
\nabla_\mu\partial^\mu -
\frac{1}{2} P_{\mu~\,\lambda}^{~\mu} \partial^\lambda 
 =
\partial^\mu\nabla_\mu +
\partial^\lambda\frac{1}{2} P_{\mu~\,\lambda}^{~\mu}  \,,
\end{equation}
and so the rules $ \partial^A|0_p\rangle= \langle 0_p|\partial^A=0$
can be exploited as in the flat case ($|0_p\rangle$ being the
unit wavefunction in $p$ space, see footnote \ref{foot:4}).

Because the covariant symbol is multiplicative with respect to $x$ (to
be shown subsequently), one could, loosely speaking, replace $\langle
x_0|\overline{f}|0\rangle$ with $\overline{f}(x_0)$, interpreted as an
operator valued function at $x_0$. In addition, in the absence of
derivatives, whether $p_\mu$ is constant or not is no longer
relevant, and one can formally integrate over $p_\mu$ instead of
$p_A$, the Jacobian implying the replacement of $g^{(\xi)}(x_0)$ with
$g(x_0)$.
\begin{eqnarray}
\langle x_0| \hat{f}|x_0\rangle
&=&
\frac{1}{\sqrt{g(x_0)}}
\int \frac{d^dp_\mu}{(2\pi)^d} \, \overline{f}(x_0)
\,.
\label{eq:4.5a}
\end{eqnarray}
In case of ambiguity the expression (\ref{eq:4.5}) should be
used. (See also Section \ref{sec:6} for further details.)

The properties of the covariant symbol are 
\begin{itemize}
\item[i)] It is a representation (an algebra homomorphism). This
follows from being defined as a similarity transformation. Actually,
the definition in (\ref{eq:4.3}) is a similarity transformation in a
extended sense, since, in general the $\nabla_\mu$'s in the formula
will fall in different representations (different rank world
tensors). In any case, whenever $\nabla_\mu$ acts on a field, it
selects, by convention, the appropriate connections corresponding to
the gauge and world representation of the field, in such a way that
the homomorphism property holds, that is\footnote{The symbols $X$ and $Y$ are
used to represent arbitrary operators. In particular $X$ is unrelated
to the vector field $X^\mu$.}
\begin{equation}
\lambda X+\mu Y\mapsto \lambda \overline{X}+\mu \overline{Y}\,,
\quad
XY\mapsto \overline{X}\overline{Y} \,.
\end{equation}
\item[ii)] The covariant symbol is a multiplicative operator (with
respect to $x$), i.e., it falls in the class $ {\cal
C}(\underline{\nabla})$, provided the original operator does not act
in $p$ space, or more precisely $\hat{f}\in {\cal C}(\underline{p})$
(it may contain $\partial^\mu$). This can be as follows: the
covariant symbol being multiplicative is equivalent to
\begin{equation}
e^{-aX}\overline{f}e^{aX}=\overline{f}
\,,\quad
\forall a_A\in 
{\cal C}(\underline{x},\underline{\nabla},\underline{Z},\underline{W},\underline{I},\underline{p},
\underline{\partial})
\end{equation}
where $aX=a_\mu X^\mu$, with $a_\mu= t^A_\mu a_A$, $a_A$ being
an arbitrary constant c-number quantity. Then
\begin{eqnarray}
&&
e^{-aX}\overline{f}e^{aX}
=e^{-aX}e^{-\frac{1}{2}\{\nabla,\partial\}}e^{-pX}
\hat{f}
e^{pX} 
e^{\frac{1}{2}\{\nabla,\partial\}}
e^{aX}
\nonumber \\
&& =
e^{-\frac{1}{2}\{\nabla,\partial\}}e^{-a\partial}e^{-aX}e^{-pX}
\hat{f}
e^{pX} 
e^{aX}
e^{a\partial}
e^{\frac{1}{2}\{\nabla,\partial\}}
\nonumber \\
&& =
e^{-\frac{1}{2}\{\nabla,\partial\}}e^{-pX}
e^{-a\partial}
\hat{f}
e^{a\partial}
e^{pX} 
e^{\frac{1}{2}\{\nabla,\partial\}}
\nonumber \\
&& =
\overline{f}, 
\qquad \hat{f}\in {\cal C}(\underline{p})
\end{eqnarray}
In the second equality we use $[\nabla_\mu,aX]=a_\mu$ to move $e^{aX}$
to the left, generating a factor $e^{a\partial}$. Since $a_\mu\partial^\mu=
a_A\partial^A$, this factor is a constant c-number and commutes with
everything except $p_A$. In the last equality we use that $\hat f$
does not contain $p_A$.
\item[iii)] It is gauge and world covariant. This follows from
using a covariant coordinate system such as Riemann normal coordinates at
$x_0$. Because the covariant symbol is multiplicative (all $x$
derivatives have already been taken) no quantities at points different
from $x_0$ are needed.
\item[iv)] It preserves the hermiticity properties of the original
operator. Assuming  hermiticity rules in $x$ space of the type
\begin{eqnarray}
&&(\lambda X+\mu Y)^\dagger =\lambda^* X^\dagger +\mu^* Y^\dagger
\,,\quad
(XY)^\dagger =Y^\dagger X^\dagger
\,,
\nonumber \\
&&
(x^\mu)^\dagger=+x^\mu
\,,\quad
(\nabla_\mu)^\dagger=-\nabla_\mu
\,,\quad
(Z_{\mu\nu})^\dagger=-Z_{\mu\nu}
\,,\quad
(R_{\mu\nu~\alpha}^{~~\,\,\beta~})^\dagger= +R_{\mu\nu~\alpha}^{~~\,\,\beta~}
\,,\quad
(T_{\mu\nu}{}^\lambda)^\dagger= +T_{\mu\nu}{}^\lambda
\,,
\end{eqnarray}
etc,
the Hermitian character of an operator is shared by its covariant
symbol by adding the prescriptions
\begin{equation}
(p_\mu)^\dagger=-p_\mu
\,,\quad
(\partial^\mu)^\dagger=+\partial^\mu
\end{equation}
(recall that we are using a purely imaginary momentum variable throughout).
\end{itemize}

Of course, in practice the calculations implied in the definition of
the covariant symbol cannot be carried out explicitly in full, a
statement that also holds for the matrix element $\langle
x|\hat{f}|x\rangle$ itself. A suitable approach compatible with its
definition, is to carry out a covariant derivative expansion of the
symbol. In this counting each $\nabla_\mu$ counts as first order, the
torsion $T_{\mu\nu}{}^\lambda$ is also first order,
$R_{\mu\nu~\alpha}^{~~\,\,\beta~}$ is second order and so on, $p_\mu$
and $\partial^\mu$ count as zeroth order.

A systematic computation is presented in next Section. By way
of illustration we show here the covariant symbol of $M$ and $\nabla_\mu$ to
second order in the derivative expansion for
a general connection and general reference
coordinate system. The result is expressed in terms of the tensors
$P_{\mu_1\mu_2\cdots\mu_n}\!{}^\alpha_{~\,\beta}$,
which generalize that in (\ref{eq:3.5}) and (\ref{eq:3.6})
\begin{equation}
[\nabla_{\mu_1},[\nabla_{\mu_2},\cdots,[\nabla_{\mu_n},p_\beta]\cdots]]
=
-P_{\mu_1\mu_2\cdots\mu_n}\!{}^\alpha_{~\,\beta}p_\alpha \,.
\label{eq:4.13}
\end{equation}
This gives
\begin{eqnarray}
\overline{M} &=& M-M_\mu\partial^\mu+\frac{1}{2}M_{\mu\nu}\partial^\mu\partial^\nu
+\frac{1}{2}M_\mu P_{\nu~\,\lambda}^{~\,\mu}\partial^\lambda\partial^\nu
+{\cal O}(\nabla^3) \,,
\nonumber \\
\overline{\nabla}_\mu &=&
p_\mu
+\frac{1}{2}P_{\alpha~\,\mu}^{~\lambda}\{p_\lambda,\partial^\alpha\} 
-\frac{1}{4}\{Z_{\alpha\mu},\partial^\alpha\}
-\frac{1}{4}(P_{\alpha\beta~\,\mu}^{~~~\lambda}
+P_{\alpha~\,\beta}^{~\,\sigma} P_{\sigma~\,\mu}^{~\lambda} )
\{p_\lambda,\partial^\alpha\partial^\beta\}
+{\cal O}(\nabla^3) \,.
\end{eqnarray}
We have used the property $P_{\mu~\,\nu}^{~\alpha} -P_{\nu~\,\mu}^{~\alpha} =
T_{\mu\nu}^{~~\,\alpha}$.  As we can see $\overline{\nabla}_\mu$ is
multiplicative in $x$ space.

We can particularize these formulas to the case of normal coordinates
at $x_0$ as RCS, but arbitrary connection. Using the results of the
next section (cf. (\ref{eq:5.16a}) and (\ref{eq:5.22a})), one obtains
at $x_0$\footnote{We recall that only for multiplicative operators can
one meaningfully take $x=x_0$.}
\begin{eqnarray}
\overline{M} &=& M-M_\mu\partial^\mu+\frac{1}{2}M_{\mu\nu}\partial^\mu\partial^\nu
+{\cal O}(\nabla^3) \,,
\label{eq:4.14}
\\
\overline{\nabla}_\mu &=&
p_\mu
+\frac{1}{4} T_{\alpha\mu}{}^\lambda \{p_\lambda,\partial^\alpha\}
-\frac{1}{4}\{Z_{\alpha\mu},\partial^\alpha\}
+\left(\frac{1}{12}
R_{\mu\alpha~\,\beta}^{~~~\lambda}
-\frac{1}{6}T_{\alpha\beta\mu}{}^\lambda
+\frac{1}{24}T_{\mu\alpha}{}^\sigma T_{\sigma\beta}{}^\lambda
\right)
\{p_\lambda,\partial^\alpha\partial^\beta\}
+{\cal O}(\nabla^3) \,.
\nonumber 
\end{eqnarray}
This result is manifestly covariant and all operators involved are
multiplicative with respect to $x$. The Hermitian properties are
explicit as well. Also note that $\overline{M}$ and
$\overline{\nabla}_\mu$ are still operators with respect to the gauge
and world indices. For instance, $Z_{\mu\nu}$ acts on world indices
yielding the curvature tensor. It is also noteworthy that this
formulas are formally independent of the domain of the operators
involved. In particular, depending on the internal and world
representation of the states, $Z_{\mu\nu}$ will act in a way or
another. E.g., on a state which is gauge singlet and world scalar
$|\phi\rangle$
\begin{equation}
 Z_{\mu\nu}|\phi\rangle=0 \,,
\end{equation}
yet
\begin{equation}
 Z_{\mu\nu}p_\lambda|\phi\rangle=
[Z_{\mu\nu},p_\lambda]|\phi\rangle+
p_\lambda Z_{\mu\nu}|\phi\rangle=
-R_{\mu\nu~\,\lambda}^{~~~\sigma}\,p_\sigma |\phi\rangle \,.
\end{equation}
When working with operators in ${\cal C}(Z)$, like $Z_{\mu\nu}$, one
should be aware of seemingly paradoxical results. E.g., for the same
state $|\phi\rangle$ as before, and for $B_\lambda$ a gauge singlet
field,
\begin{equation}
0=\langle \phi|Z_{\mu\nu}B_\lambda |\phi\rangle
=-\langle \phi|R_{\mu\nu~\,\lambda}^{~~~\sigma}\,B_\sigma |\phi\rangle
\end{equation}
(using $\langle\phi|Z_{\mu\nu}= Z_{\mu\nu}|\phi\rangle=0$). Indeed the
result is zero, since the state
$R_{\mu\nu~\,\lambda}^{~~~\sigma}\,B_\sigma |\phi\rangle$ is a rank
three tensor and cannot connect with the scalar state
$|\phi\rangle$. There is also no contradiction if one uses instead the
operator $Z_{\mu\nu}B^\mu C^\nu$ (again $B^\mu$, $C^\nu$ gauge
singlets) which is a scalar (and the previous argument would not
apply), since in this case the operator $[Z_{\mu\nu},B^\mu C^\nu]$
itself vanishes.

A preliminary definition of the covariant symbol was given in
(\ref{eq:4.3}). As final definition we take (\ref{eq:4.3}) but using
as RCS the Riemann normal coordinates associated to the given connection
(these coordinates are defined for any connection), at each point $x$.
That is, we use a different RCS at each point. This is perfectly well
defined since the operator is multiplicative and so equivalent to a
function. For the same reason, the algebra homomorphism property is
also not spoiled. There is an ambiguity in that normal coordinates at
$x$ are unique modulo a rigid general linear transformation. However,
such ambiguity does not reflect on the form of the covariant symbol
when written in terms of $p_\mu$ and $\partial^\mu$.

\section{Computation of the covariant symbols}
\label{sec:5}

In this Section we proceed to set up a systematic computation of the
covariant symbols within a derivative expansion. The expansion is taken
to fourth order.

\subsection{Arbitrary reference coordinate system and arbitrary connection}

Momentarily, we will work with an arbitrary RCS and arbitrary
connection. The quantities $t^A_\mu$, $t^\mu_A$, $p_A$, $p_\mu$,
$\partial^A$ and $\partial^\mu$ have already been defined and some of
their properties noted in previous sections. Here we only note the
equivalent definition
\begin{equation}
t^A_\mu:=[\nabla_\mu,\xi^A]\,,
\end{equation}
where $\xi^A$ is regarded as a c-number multiplicative operator. Next
we introduce the world scalar operator (in ${\cal
C}(\underline{W},\underline{I},\underline{p},\underline{\partial})$)
\begin{equation}
\nabla_A:=\frac{1}{2}\{\nabla_\mu,t^\mu_A\} \,, 
\label{eq:5.5}
\end{equation}
which has the property
\begin{equation}
\nabla_\mu=\frac{1}{2}\{\nabla_A,t_\mu^A\} \,,
\label{eq:5.6}
\end{equation}
as is readily shown. For this and similar manipulations the following
lemma is useful

{\bf Lemma:} If the set of operators $A_i$ satisfies
\begin{equation}
[A_i,A_j]=[[B,A_i],A_j]=0 \quad\text{for all~} i,j
\end{equation}
then
\begin{equation}
\frac{1}{4}\{\{B,A_i\},A_j\}=
\frac{1}{2}\{B,A_iA_j\}
\quad\text{for all~} i,j\,.
\end{equation}
Applying the lemma, (\ref{eq:5.6}) follows from the definition
(\ref{eq:5.5}) and the properties
$[t_\mu^A,t_B^\mu]=[[\nabla_\mu,t_\nu^A],t^\alpha_B]=0$. A crucial
property of $\nabla_A$ is that $[\nabla_A,X]$ is multiplicative
provided $X$ is multiplicative, as is easily shown. As note before,
this property is not enjoyed by $\nabla_\mu$: to have that the
stronger assumption $X\in {\cal C}(\underline{\nabla},\underline{Z})$
is needed.\footnote{If $\phi$ is a c-number multiplicative operator
and $X$ is merely multiplicative,
$[[\nabla_A,X],\phi]=[[\nabla_A,\phi],X]=0$ since $[\nabla_A,\phi]$ is
again a c-number multiplicative operator, however, $[\nabla_\mu,\phi]$
is not (it contains a world index).}

In the RCS it holds $p_\mu X^\mu=p_A \xi^A$, and also
$\frac{1}{2}\{\nabla_\mu,\partial^\mu\}=\nabla_A\partial^A$ (using
(\ref{eq:5.5}) and the fact that $\partial^A$ commutes $\nabla_\mu$
and $t^\mu_A$). Therefore, we can reexpress the definition
(\ref{eq:4.3}) of the covariant symbol in the form
\begin{equation}
\overline{f}=e^{-\partial^A \nabla_A}e^{-p_B \xi^B}\hat{f}\, e^{p_C\xi^C}
e^{\partial^D\nabla_D} \,.
\label{eq:5.9}
\end{equation}

The commutation properties of the quantities $\xi^A$, $\nabla_A$, $p_A$
and $\partial^A$ are as follows
\begin{eqnarray}
&&[\xi^A,\xi^B]=[\xi^A,p_B]=[\xi^A,\partial_B]=
[\nabla_A,p_B]=[\nabla_A,\partial^B]=
[p_A,p_B]=[\partial^A,\partial^B]=
0 \,,
\nonumber \\
&&
[\nabla_A,\xi^B]=\delta^B_A \,,
\quad
[\partial^A,p_B]=\delta^A_B
\,,
\quad
[\nabla_A,\nabla_B]:=Z_{AB}\,.
\end{eqnarray}
We have introduced the operator $Z_{AB}$. Recursively we find
\begin{eqnarray}
&&Z_{A_1\cdots A_n}:=[\nabla_{A_1},Z_{A_2\cdots A_n}]
 \,,
\nonumber \\
&&
[Z_{A_1\cdots A_n},\xi^B]=
[Z_{A_1\cdots A_n},,p_B]=
[Z_{A_1\cdots A_n},\partial^B]=0
\,.
\end{eqnarray}
In particular, note that the operators $Z_{A_1\cdots A_n}$ are
multiplicative since they commute with $\xi^A$.

The important observation is that the commutation relations of
$\xi^A$, $\nabla_A$, $p_A$ and $\partial^A$ are identical to those of
the flat case, as is also the definition of the covariant
symbol in terms of these operators, (\ref{eq:5.9}). This immediately
implies that the analogous of (\ref{eq:2.12}) hold
\begin{eqnarray}
{\overline{M}} &=& 
M -M_S\,\partial^S
+\frac{1}{2!}M_{SS}\,(\partial^S)^2
-\frac{1}{3!}M_{SSS}\,(\partial^S)^3
+\cdots \,,
\quad
M\in
 {\cal C}(\underline{\nabla},\underline{Z},\underline{p},
\underline{\partial})
 \nonumber \\ 
{\overline{\nabla}_A} &=& 
 p_A
-\frac{1}{2!}Z_{SA}\,\partial^S
+\frac{2}{3!}Z_{SSA}\,(\partial^S)^2
-\frac{3}{4!}Z_{SSSA}\,(\partial^S)^3
+\cdots \,.
\label{eq:5.12}
\end{eqnarray}
with $M_{A}=[\nabla_A,M]$, $M_{AB}=[\nabla_A,[\nabla_B,M]]$, and so
on, and $S$ standing for contracted symmetrized indices of the type
$A$, $B$, $\ldots$, (cf. Appendix \ref{app:A}). (Actually, the
equation for ${\overline{M}}$ holds too for $M\in {\cal
C}(\underline{\nabla},\underline{p}, \underline{\partial})$.)

Unfortunately, this simple result is not sufficient. We need
${\overline{\nabla}_\mu}$ instead of ${\overline{\nabla}_A}$, and
quantities formed with $M$ and $\nabla_\mu$ instead of $\nabla_A$ and
$Z_{A_1,A_2,\ldots,A_n}$.

\subsection{Riemann normal coordinates and arbitrary connection}

Let us consider first ${\overline{M}}$. Since $M\in {\cal C}(\underline{Z})$,
$M$ commutes with $t^\mu_A$ and $t_\mu^A$ and their derivatives. Hence, we find
\begin{eqnarray}
M_A &=& M_\mu \,t^\mu_A \,, \nonumber \\
M_{AB} &=& M_{\mu\nu}\,t^\mu_A \, t^\nu_B + M_\nu \, t^\nu_{\mu B}\, t^\mu_A \,,
\end{eqnarray}
and so on. As before, we use the notation
\begin{equation}
t_{\mu_1\ldots\mu_n A}^\lambda =[ \nabla_{\mu_1}, t_{\mu_2\ldots\mu_n A}^\lambda] \,,
\quad
t_{\mu_1\ldots\mu_n \lambda}^A =[ \nabla_{\mu_1}, t_{\mu_2\ldots\mu_n \lambda}^A] \,.
\end{equation}
The derivatives of the type $t_{\mu_1\ldots\mu_n A}^\lambda$ can be expressed in terms of 
$t_B^\lambda$ and $t_{\mu_1\ldots\mu_n \lambda}^B$ using the relations (\ref{eq:5.2}), e.g.
\begin{equation}
t^\nu_{\mu A}= -t^\nu_B\, t^\lambda_A\, t^B_{\mu\lambda} \,.
\end{equation}
This gives for the term with two derivatives in ${\overline{M}}$
\begin{equation}
M_{SS}= (M_{ss} + M_\lambda \, t^A_{ss}\,t^\lambda_A) t^s_S \, t^s_S \,.
\label{eq:5.16}
\end{equation}

To proceed further we make a choice of RCS (for given base point
$x_0$), namely, we choose the usual Riemann normal coordinates at
$x_0$. These are the coordinates such that the curves $\xi^A(t)=t v^A$
are geodesics passing through $x_0$, the geodesics being the
straightest lines with respect to the given connection. A practical
equivalent definition is to take the coordinates $\xi^A$ so that
\begin{equation}
\xi^A(x_0)=0\,, \qquad
t^A\!\!\!\!\underbrace{{}_{ss\ldots s}}_n\Big|_{x_0} =0 \qquad  \text{for~~} n \ge 2 \,.
\label{eq:5.17}
\end{equation}
(This means that, for $n\ge 2$, the completely symmetrized component
of $t_{\mu_1\ldots\mu_n}^A$ vanishes.)  In this form the normal
coordinates system was used in a similar context in
\cite{Gusynin:1990bu}. Given $x_0$ and $t^A_\mu$ at $x_0$, the normal
coordinates are locally unique, since the $t_{\mu_1\ldots\mu_n
\lambda}^A$ at $x_0$ can be obtained recursively in terms of the
curvature and torsion tensors and their derivatives, using the
definition above. For instance,
\begin{equation}
t^A_{\mu\nu}=\nabla_\mu\nabla_\nu\xi^A =
t^A_{\nu\mu}- T_{\mu\nu}^{~~\lambda}\,t^A_\lambda
\label{eq:5.18}
\end{equation}
and using (\ref{eq:5.17}) (i.e., $t^A_{\mu\nu}$ is purely antisymmetric at $x_0$)
\begin{equation}
t^A_{\mu\nu}= 
-\frac{1}{2} T_{\mu\nu}^{~~\lambda}\,t^A_\lambda \,, \qquad \text{at~} x_0 \,.
\label{eq:5.16a}
\end{equation}
Likewise, starting from
\begin{equation}
t^A_{\mu\nu\alpha}+\text{five permutations} =0\,, \qquad \text{at~} x_0 \,,
\end{equation}
and using the identity
\begin{equation}
t^A_{\alpha\beta\gamma}=
t^A_{\beta\alpha\gamma} 
-R_{\alpha\beta~\gamma}^{~~\,\,\lambda~} \, t^A_{\lambda}
-T_{\alpha\beta}^{~~\,\lambda}\,t^A_{\lambda\gamma} 
\end{equation}
 (plus derivatives of (\ref{eq:5.18})) to bring the five permutations
to coincide with the first ordering, gives $t^A_{\mu\nu\alpha}$ at
$x_0$. This yields
\begin{eqnarray}
t^A_{\alpha\beta\gamma} &=& 
\Big(
-\frac{1}{3}R_{\alpha\beta}{}^\lambda\!{}_\gamma
+\frac{1}{3}R_{\gamma\alpha}{}^\lambda\!{}_\beta
-\frac{1}{2}T_{\alpha\beta\gamma}{}^\lambda
+\frac{1}{6}T_{\beta\gamma\alpha}{}^\lambda
-\frac{1}{6}T_{\gamma\alpha\beta}{}^\lambda
\nonumber \\ && \qquad
-\frac{1}{4} T_{\beta\gamma}{}^\sigma T_{\sigma \alpha}{}^\lambda
-\frac{1}{12}T_{\gamma\alpha}{}^\sigma T_{\sigma \beta}{}^\lambda
+\frac{1}{12}T_{\alpha\beta}{}^\sigma T_{\sigma \gamma}{}^\lambda
\Big)
 t^A_{\lambda}
\,, \qquad \text{at~} x_0 \,.
\label{eq:5.22a}
\end{eqnarray}
The same technique applies for computing any higher derivative of $\xi^A$ at
$x_0$.

Let us come back now to the evaluation of $M_{SS}$. Clearly for normal
coordinates (but arbitrary connection), (\ref{eq:5.16}) reduces to
$M_{SS}= M_{ss} t^s_S \, t^s_S$ at the origin. In fact a similar
reduction happens to all orders, that is,\footnote{This can be seen
recursively as follows. $M_{S\ldots S}$ contains a first term
$M_{s\ldots s} \, t^s_S \cdots t^s_S$ plus other terms containing a
factor $t^A_{s\ldots s}$ (with two or more $s$'s). Now, in the present
case (namely, $M\in {\cal
C}(\underline{\nabla},\underline{Z},\underline{p},
\underline{\partial})$ but not, e.g., for $Z_{\mu\nu}$), each new
$\nabla_S$ is equivalent to $t^s_S\nabla_s$. If $\nabla_s$ acts on a
factor $t^s_S$ in the first term, the formula
$$
t^\lambda_{sS}= -t_A^\lambda \, t^s_S \, t^A_{ss}
$$ applies and this vanishes at $x_0$. Likewise, if $\nabla_s$ acts on
the other terms, there will always remain a $t^A_{s\ldots s}$ factor
with two or more $s$'s. So the only surviving piece at $x_0$ is that
obtained by $\nabla_s$ acting on $M_{s\ldots s}$. This yields
(\ref{eq:5.24}). See footnote \ref{foot:19} for an alternative proof.}
\begin{equation}
M_{S\ldots S}= M_{s\ldots s} \, t^s_S \cdots t^s_S \,, \qquad
\text{at~} x_0 \,.
\label{eq:5.24}
\end{equation}

In summary, the covariant symbol of $M$, to all orders in the
derivative expansion, is given by a expression fully analogous to that
of the purely gauge case, namely,
\begin{eqnarray}
{\overline{M}} &=& 
M - M_s\,\partial^s
+\frac{1}{2!}M_{ss}\,(\partial^s)^2
-\frac{1}{3!}M_{sss}\,(\partial^s)^3
+\cdots \,,
\quad
M\in
 {\cal C}(\underline{\nabla},\underline{Z},\underline{p},
\underline{\partial})
\label{eq:5.26}
\end{eqnarray}
where $s$ are symmetrized world symbols. Note that, unlike
(\ref{eq:5.12}), this expression does not apply for arbitrary
operators in ${\cal C}(\underline{\nabla},\underline{p},
\underline{\partial})$ such as $Z_{\mu\nu}$.

Next, we need to introduce world-index counterparts of $Z_{A_1\ldots A_n}$.
As discussed in section \ref{sec:3}, the objects
$[\nabla_{\mu_1},[\cdots,\nabla_{\mu_n}]\cdots]$ are not in general
multiplicative operators. Instead, we recursively define
\begin{eqnarray}
Z_{\mu\nu} &=& [\nabla_\mu,\nabla_\nu]
+\frac{1}{2}\{\nabla_\lambda,T_{\mu\nu}^{~~\,\lambda}\}
 \,, \nonumber \\
Z_{\alpha\mu\nu} &=& [\nabla_\alpha,Z_{\mu\nu}]
-\frac{1}{2}\{\nabla_\lambda , R^{~~\,\lambda}_{\mu\nu~\alpha} \}  \,,
 \nonumber 
\\
&\vdots& 
\label{eq:Zs}
 \\
Z_{\alpha\mu_1\cdots\mu_n} &=& [\nabla_\alpha,Z_{\mu_1\cdots\mu_n}]
-\frac{1}{2}\{\nabla_\lambda , R^{~~~~~~~\lambda}_{\mu_1\cdots\mu_n~\alpha} \} \,, 
 \nonumber
\end{eqnarray}
(with $R_{\sigma\mu\nu~\beta}^{~~~\,\alpha}:=\nabla_\sigma
R_{\mu\nu~\beta}^{~~\,\alpha}$, etc). These operators are
multiplicative, and indeed for a gauge singlet $V^\sigma$ they satisfy
\begin{eqnarray}
[Z_{\mu\nu},V^\sigma]  &=& R_{\mu\nu~\lambda}^{~~\,\,\sigma}\,V^\lambda \,,
\nonumber \\
~ [ Z_{\alpha\mu\nu},V^\sigma]   &=& 
R_{\alpha\mu\nu~\lambda}^{~~~~\,\sigma}\,V^\lambda  
\,, \nonumber 
\\
&\vdots& 
\qquad\qquad\qquad\qquad 
V\in{\cal C}(\underline{\nabla},\underline{Z},\underline{I})
\label{eq:5.30}
\\
~  [Z_{\mu_1\cdots\mu_n},V^\sigma]  &=& 
R_{\mu_1\cdots\mu_n ~\lambda}^{~~~~~~~\,\sigma}\,V^\lambda
\nonumber
\end{eqnarray}
(and of course a similar action on each world index in the case of
tensors).  In addition they are antihermitian. In terms of these, one
obtains the following relations at $x_0$,\footnote{Naturally,
instead of (\ref{eq:Zs}), we could have adopted a definition of the type
$$
Z^\prime_{\mu_1\cdots\mu_n} =
\frac{1}{2}\{ t^{A_1}_{\mu_1}\cdots t^{A_n}_{\mu_n}  , Z_{{A_1}\cdots {A_n}} \} ,
$$
which has all the good properties, and in particular
$$
Z_{{A_1}\cdots {A_n}} =
\frac{1}{2}\{ t_{A_1}^{\mu_1}\cdots t_{A_n}^{\mu_n}  , Z^\prime_{\mu_1\cdots\mu_n} \} \,.
$$ However, the relations similar to (\ref{eq:5.30}) become more
complicated. The definition adopted corresponds to
$Z_{\alpha\mu_1\ldots\mu_n} = \frac{1}{2}\{ t^A_\alpha, [\nabla_A
,Z_{\mu_1\ldots\mu_n} ] \}$.}
\begin{eqnarray}
  Z_{AB} &=& \frac{1}{2}\{t_A^\alpha \, t_B^\beta , Z_{\alpha\beta} \} \,,
\qquad\qquad\qquad\qquad\qquad\qquad (\text{all at~} x_0)
\nonumber 
\\
Z_{ABC} &=& \frac{1}{2} \big\{t_A^\alpha \, t_B^\beta \, t_C^\mu ,
 Z_{\alpha\beta\mu} 
+\frac{1}{4}\{Z_{\beta\lambda} , T_{\alpha\mu}{}^\lambda\}
-\frac{1}{4}\{Z_{\mu\lambda} , T_{\alpha\beta}{}^\lambda\}
\big\}
\,.
\label{eq:5.24a}
\end{eqnarray}

We can proceed now to the evaluation of $\overline{\nabla}_\mu$. To do this
we use the relation
\begin{equation}
\overline{\nabla}_\mu=\frac{1}{2}\{\overline{\nabla}_A,\overline{t}_\mu^A\}
\label{eq:5.25}
\end{equation}
which follows from (\ref{eq:5.6}) and the homomorphism property of the
covariant symbols. The quantity $\overline{t}_\mu^A$ is easily
obtained two second order using $M=t^A_\mu$ in (\ref{eq:5.26}) and the
formulas (\ref{eq:5.16a}) and (\ref{eq:5.22a})
\begin{eqnarray}
\overline{t}_\mu^A &=& t^A_\lambda \left[
\delta^\lambda_\mu 
+\frac{1}{2} T_{s\mu}{}^\lambda\,\partial^s
+ \left(
 \frac{1}{6}R_{\mu s ~ s}^{~~\,\lambda}
-\frac{1}{3}T_{ss\mu}{}^\lambda 
+\frac{1}{12}T_{s\mu}{}^\sigma T_{s \sigma }{}^\lambda 
\right)
(\partial^s)^2
+{\cal O}(\nabla^3)
\right] \,.
\end{eqnarray}
On the other hand $\overline{\nabla}_A$ is obtained to second order
from (\ref{eq:5.12}) and the first equation in (\ref{eq:5.24a}). In
this way we reproduce the result for $\overline{\nabla}_\mu$ in
(\ref{eq:4.14}).

\subsection {Riemannian connection results}

From now on we restrict ourselves to the Riemannian connection, since
the absence of torsion considerably simplifies the expressions.

For the Riemannian connection one finds at $x_0$
\begin{eqnarray}
t_{\mu\nu}^A &=& 0 \,, \nonumber \\
t^A_{\alpha\mu\nu} &=& \frac{1}{3}\left(
R^\lambda_{~\mu\nu\alpha}+ R^\lambda_{~\nu\mu\alpha}\right)
 t^A_{\lambda}
\,, \qquad \text{at~} x_0 \,.
\label{eq:5.22b}
\end{eqnarray}
Some higher order results needed to obtain $\overline{t}_\mu^A$ are as
follows
\begin{eqnarray}
t_{s\mu}^A &=& 0 \,, \nonumber \\
t_{ss\mu}^A &=& \frac{1}{3}R_{\mu s ~ s}^{~~\,\lambda}\,t^A_\lambda
\,,  \nonumber \\
t_{sss\mu}^A &=& \frac{1}{2}R_{s\mu s ~ s}^{~~~\,\lambda}\,t^A_\lambda
\,, 
  \label{eq:5.23} \\
t_{ssss\mu}^A &=& 
\left[\frac{3}{5}R_{ss\mu s ~ s}^{~~~~\,\lambda}
+ \frac{7}{15} R_{\mu s ~ s}^{~~\,\sigma}R_{\sigma s ~ s}^{~~\,\lambda}
\right]t^A_\lambda
\,, 
\qquad \text{all at~} x_0 \,.
\nonumber
\end{eqnarray}
With (\ref{eq:5.26}) this gives
\begin{eqnarray}
\overline{t}_\mu^A &=& t^A_\lambda \left[
\delta^\lambda_\mu 
+ \frac{1}{6}R_{\mu s ~ s}^{~~\,\lambda}\,(\partial^s)^2
- \frac{1}{12}R_{s\mu s ~ s}^{~~~\,\lambda}\,(\partial^s)^3
\right. \nonumber \\ 
&&  \quad \left.
+ \left(\frac{1}{40}R_{ss\mu s ~ s}^{~~~~\,\lambda}
+ \frac{7}{360} R_{\mu s ~ s}^{~~\,\sigma}R_{\sigma s ~ s}^{~~\,\lambda}
\right)(\partial^s)^4
+{\cal O}(\nabla^5)
\right] \,.
\end{eqnarray}

In addition, for the Riemannian connection one has the Bianchi identities
\begin{eqnarray}
&& 0=  
Z_{\alpha_1\cdots\alpha_n\mu_1\mu_2\mu_3}
+ Z_{\alpha_1\cdots\alpha_n\mu_2\mu_3\mu_1}+Z_{\alpha_1\cdots\alpha_n\mu_3\mu_1\mu_2}\,,
\end{eqnarray}
and the following relations (which hold at $x_0$)
\begin{eqnarray}
  Z_{AB} &=& \frac{1}{2}\{t_A^\alpha \, t_B^\beta , Z_{\alpha\beta} \} \,,
\nonumber 
\\
Z_{ABC} &=& \frac{1}{2}\{t_A^\alpha \, t_B^\beta \, t_C^\mu ,
 Z_{\alpha\beta\mu} \} \,, 
\qquad\qquad\qquad\qquad\qquad\qquad (\text{all at~} x_0)
\\
Z_{ABCD} &=& \frac{1}{2}\big\{
t_A^\alpha \, t_B^\beta \, t_C^\mu \, t_D^\nu ,
Z_{\alpha\beta\mu\nu}
+\frac{1}{6}\{Z_{\lambda\mu},
R^\lambda_{~\beta\nu\alpha}+R^\lambda_{~\nu\beta\alpha}\}
-\frac{1}{6}\{Z_{\lambda\nu},
R^\lambda_{~\beta\mu\alpha}+R^\lambda_{~\mu\beta\alpha}\}
\big\} 
\,.
\nonumber
\end{eqnarray}

Combining all the previous relations and (\ref{eq:5.25}) we can now
write down the expression for $\overline{\nabla}_\mu$, which we have
computed to four derivatives
\begin{eqnarray}
\overline{\nabla}_\mu &=&
\overline{\nabla}_\mu{}^{(0)} 
+\overline{\nabla}_\mu{}^{(1)} 
+\overline{\nabla}_\mu{}^{(2)}
+\overline{\nabla}_\mu{}^{(3)}
+\overline{\nabla}_\mu{}^{(4)}   
+{\cal O}(\nabla^5)
\end{eqnarray}
The result is as follows
\begin{eqnarray}
\overline{\nabla}_\mu{}^{(0)} &=& p_\mu 
\,,
\nonumber \\
\overline{\nabla}_\mu{}^{(1)} &=&  0
\,,
\nonumber \\
\overline{\nabla}_\mu{}^{(2)} &=& 
-\frac{1}{4}\{ Z_{s\mu},\partial^s\}
+\frac{1}{12}\{ [ Z_{s\mu},p_s],(\partial^s)^2\}
\,,
\nonumber \\
\overline{\nabla}_\mu{}^{(3)} &=& 
\frac{1}{6}\{ Z_{ss\mu},(\partial^s)^2\}
-\frac{1}{24}\{ [ Z_{ss\mu},p_s],(\partial^s)^3\}
\,,
\nonumber \\
\overline{\nabla}_\mu{}^{(4)} &=& 
-\frac{1}{16}\{ Z_{sss\mu},(\partial^s)^3 \}
+ \frac{1}{80}\{ [Z_{sss\mu},p_s],(\partial^s)^4 \}
+ \frac{1}{48}\{ Z_{s\lambda}, [ Z_{s\mu},\partial^\lambda](\partial^s)^2 \}
\nonumber \\
&&
-\frac{7}{720}\{ [Z_{s\lambda},p_s], [Z_{s\mu},\partial^\lambda](\partial^s)^3 \} \,.
\end{eqnarray}
The result has been written in a manifestly antihermitian form.
Although the Riemann tensor does not appear, these formulas hold only
for the Riemann connection. The formulas have been verified in various
ways. In particular, for $M\in {\cal
C}(\underline{\nabla},\underline{Z},\underline{p},
\underline{\partial}) $ one can apply the expansion (\ref{eq:5.26}) to
$M$ and also to $M_\mu=[\nabla_\mu,M]$ since it fall in the same
class. Then one can check that the formulas preserve the homomorphism
property, in the form
\begin{equation}
\overline{M}_\mu= [\overline{\nabla}_\mu,\overline{M}] \,.
\end{equation} 

Another check comes from computing the covariant symbol of
$Z_{\mu\nu}$ (for which (\ref{eq:5.26}) does not apply). A direct
computation to four derivatives gives
\begin{eqnarray}
\overline{Z}_{\mu\nu} &=&
Z_{\mu\nu} -\frac{1}{2}\{ Z_{s\mu\nu},\partial^s\}
+\frac{1}{4}\{ Z_{ss\mu\nu},(\partial^s)^2\}
+{\cal O}(\nabla^5) \,.
\end{eqnarray}
As we have verified, this expression satisfies
\begin{equation}
\overline{Z}_{\mu\nu}= [\overline{\nabla}_\mu,\overline{\nabla}_\nu] \,.
\end{equation}
These checks would serve to determine some of the coefficients in the
expression of $\overline{\nabla}_\mu$, but not all.

Another operator of great interest in applications is the Laplacian,
\begin{equation}
\Delta=g^{\mu\nu}\nabla_\mu\nabla_\nu \,.
\end{equation}
Since $[\nabla_\lambda,g_{\mu\nu}]=0$, an application of (\ref{eq:5.26}) gives
\begin{equation}
\overline{g}_{\mu\nu}= g_{\mu\nu} \,.
\end{equation}
(Also clear from the definition (\ref{eq:4.3}) since $g_{\mu\nu}$
commutes with all operators there.) Therefore we can use our previous
results to obtain the covariant symbol of the Laplacian, by means of
\begin{equation}
\overline{\Delta}= g^{\mu\nu}\overline{\nabla}_\mu\overline{\nabla}_\nu \,.
\end{equation} 
This yields an expansion of $\overline{\Delta}$ to four derivatives,
\begin{eqnarray}
\overline{\Delta} &=&
\overline{\Delta}{}^{(0)} 
+\overline{\Delta}{}^{(1)}
+\overline{\Delta}{}^{(2)}  
+\overline{\Delta}{}^{(3)}
+\overline{\Delta}{}^{(4)}
+{\cal O}(\nabla^5)
\end{eqnarray}
with
\begin{eqnarray}
\overline{\Delta}{}^{(0)}  &=& p_\mu p^\mu
\,,
\nonumber \\
\overline{\Delta}{}^{(1)} &=&  0
\,,
\nonumber \\
\overline{\Delta}{}^{(2)} &=&
-\frac{1}{2}\{ Z_{s\mu},p^\mu\partial^s\}
+\frac{1}{3}[[ Z_{s\mu},p^\mu],\partial^s]
+\frac{1}{6}\{[ Z_{s\mu},p_s]p^\mu,(\partial^s)^2 \}
\,,
\nonumber \\
\overline{\Delta}{}^{(3)} &=&
\frac{1}{6}\{ Z_{ss\mu}, \{p^\mu,(\partial^s)^2\}\}
-\frac{2}{3}[ Z^\mu_{~ \,s\mu},\partial^s]
-\frac{1}{12}\{[ Z_{ss\mu},p_s]p^\mu,(\partial^s)^3 \}
\,,
\nonumber \\
\overline{\Delta}{}^{(4)} &=&
-\frac{1}{16} \{ Z_{sss\mu}, \{p^\mu,(\partial^s)^3\}\}
+\frac{1}{40} \{ [ Z_{sss\mu},p_s] p^\mu,(\partial^s)^4]\}
\nonumber \\ 
&&
-\frac{1}{16} \{ Z_{s\mu}, \{ [Z_s^{~\mu},p_s],(\partial^s)^3 \} \}
+\frac{1}{8} \{ Z_{s\mu} Z_s^{~\mu}, (\partial^s)^2  \}
\nonumber \\ 
&&
+\frac{1}{30} \{ [Z_{s\mu},p_s][Z_s^{~\mu},p_s],(\partial^s)^4 \}
+\frac{1}{60}[Z^\mu_{~\,\alpha},\partial^\alpha] [Z_{\mu\beta},\partial^\beta]
\nonumber \\ 
&&
+\frac{2}{45}[Z^\mu_{~\,\alpha},\partial^\beta] [Z_{\mu\beta},\partial^\alpha]
+\frac{2}{45}[Z^\mu_{~\,\alpha},\partial^\beta] [Z_\mu^{~\alpha},\partial_\beta]
\nonumber \\ 
&&
+\frac{1}{3}[Z_{s\mu\nu}^{~~~\,\mu},\partial^\nu]\partial^s
+\frac{1}{60}[Z_{\mu s\nu}^{~~~\,\mu},\partial^\nu]\partial^s
-\frac{1}{40}[Z_{\mu~\nu s}^{~\mu},\partial^\nu]\partial^s
\,.
\end{eqnarray}
Once again the result has been written in an explicit Hermitian
form. It is noteworthy that in all terms the metric has been used once
to raise one index (as in the Laplacian itself), except in the second
term with coefficient $2/45$ in $\overline{\Delta}{}^{(4)}$, in which
the metric is used thrice. This is because specific properties of the
Riemann connection have been used in simplifying the formula. (The
metric has to appear an odd number of times since the Laplacian is odd
under $g_{\mu\nu}\to - g_{\mu\nu}$ whereas the connection itself is
even.)

The previous formulas can also be brought to a more systematic or
``standard'' form. We define such standard form by the requirement
that the quantities $R$, $Z$, $p$ and $\partial$ appear in the
expressions in this very order (i.e., the $\partial$'s occupy the
rightmost position, then the $p$'s, and so on.) Of course, in standard
form hermiticity is no longer manifest. For the covariant symbol of
the covariant derivative, we obtain
\begin{eqnarray}
\overline{\nabla}_\mu{}^{(2)} &=& 
\frac{1}{2} Z_{\mu s}\partial^s
+\frac{1}{6}\Rc_{\mu s}\,\partial^s
+\frac{1}{6} R_{\mu s ~ s}^{~~\,\alpha}\, p_\alpha (\partial^s)^2
\,,
\nonumber \\
\overline{\nabla}_\mu{}^{(3)} &=& 
\frac{1}{3}Z_{ss\mu}(\partial^s)^2
+\frac{1}{8} R_{\alpha\mu s ~ s}^{~~~\,\,\alpha} (\partial^s)^2
-\frac{1}{8} \Rc_{ss\mu} (\partial^s)^2
-\frac{1}{12}R_{s\mu s ~ s}^{~~~\,\alpha} p_\alpha (\partial^s)^3
\,,
\nonumber \\
\overline{\nabla}_\mu{}^{(4)} &=& 
-\frac{1}{8} Z_{sss\mu} (\partial^s)^3
+\frac{1}{24} R_{\mu s ~ s}^{~~\,\alpha} Z_{\alpha s} (\partial^s)^3
+\frac{1}{20}\Rc_{sss\mu} (\partial^s)^3
-\frac{1}{10} R_{s\alpha \mu s ~ s}^{~~~~~\alpha} (\partial^s)^3
\nonumber \\ &&
-\frac{13}{720} \Rc_{s\alpha} R_{\mu s ~ s}^{~~\,\alpha} (\partial^s)^3
+\frac{1}{90} R_{\alpha\mu~ s}^{~~\,\,\beta} \, R_{\beta s ~ s}^{~~\,\alpha} (\partial^s)^3
\nonumber \\ &&
+\frac{1}{40} R_{ss\mu s ~ s}^{~~~~\,\alpha} \, p_\alpha (\partial^s)^4
+\frac{7}{360} R_{\mu s ~ s}^{~~\,\alpha} \, R_{\alpha s ~ s}^{~~\,\beta} 
\, p_\beta  (\partial^s)^4
\,.
\label{eq:5.39}
\end{eqnarray}
In the same way, for the covariant symbol of the Laplacian,
\begin{eqnarray}
\overline{\Delta}{}^{(2)} &=&
\frac{1}{6}\R
+ Z^\alpha\!{}_s \, p_\alpha \partial^s
-\frac{1}{3}\Rc^\alpha\!{}_s \, p_\alpha \partial^s
+\frac{1}{3} R^\alpha\!{}_s{}^\beta\!{}_s \, p_\alpha p_\beta (\partial^s)^2
\,,
\nonumber \\
\overline{\Delta}{}^{(3)} &=&
-\frac{1}{3} Z_{\alpha ~ s}^{~\alpha} \partial^s 
-\frac{1}{4}\R_s \, \partial^s
\nonumber \\ && 
+\frac{2}{3} Z_{ss\alpha}\, p^\alpha (\partial^s)^2
+\frac{1}{6} \Rc_{ss\alpha} \, p^\alpha (\partial^s)^2
+\frac{1}{6} R_{~\alpha s ~ s}^{\alpha~~\beta}\, p_\beta (\partial^s)^2
\nonumber \\ &&
-\frac{1}{6} R_{s~ s~ s}^{~\alpha~\beta}\,p_\alpha p_\beta  (\partial^s)^3
\,,
\nonumber \\
\overline{\Delta}{}^{(4)} &=&
\frac{3}{20}\R_{ss}(\partial^s)^2
-\frac{1}{10} R_{~ \beta \alpha s ~ s}^{\alpha~~~\,\beta} (\partial^s)^2
+\frac{1}{4}Z_{s ~ \alpha s}^{~\alpha} (\partial^s)^2
\nonumber \\ &&
+\frac{29}{120}\Rc_{~ s}^\alpha \Rc_{\alpha s} (\partial^s)^2
+\frac{1}{4}\Rc_{~ s}^{\alpha} \, Z_{\alpha s} (\partial^s)^2
-\frac{31}{120}\Rc^\alpha_{~\beta} R_{\alpha s ~ s}^{~~\,\beta} (\partial^s)^2
\nonumber \\ &&
-\frac{1}{60} R_{s~~\,\gamma}^{~\alpha\beta} \,R_{s\alpha~\beta}^{~~\,\gamma} (\partial^s)^2
+\frac{1}{4}Z^\alpha_{~ \, s} \, Z_{\alpha s} (\partial^s)^2
-\frac{1}{20}\Rc_{sss\alpha}\, p^\alpha  (\partial^s)^3
\nonumber \\ &&
-\frac{3}{20} R_{s ~ \alpha s ~ s}^{~\alpha~~\beta} \, p_\beta  (\partial^s)^3
-\frac{1}{4} Z_{sss\alpha}\, p^\alpha  (\partial^s)^3
+\frac{13}{120} \Rc_{~ s}^{\alpha} \, R_{\alpha s ~ s}^{~~\,\beta} \, p_\beta (\partial^s)^3
\nonumber \\ &&
-\frac{1}{15} R_{~ s ~ s }^{\alpha \,\beta} \, 
         R_{s \alpha ~\beta}^{~~\,\gamma} \, p_\gamma (\partial^s)^3
+\frac{1}{4} R_{~ s ~ s }^{\alpha\,\beta} \, Z_{\alpha s} \, p_\beta (\partial^s)^3
\nonumber \\ &&
+\frac{1}{20} R_{ss~ s ~ \,s}^{~~\alpha~\beta} \, p_\alpha p_\beta (\partial^s)^4
+\frac{1}{15} R_{~ s ~ s}^{\alpha\,\gamma} R_{\gamma s ~ s }^{~~\,\beta} \, p_\alpha p_\beta 
(\partial^s)^4
\,.
\label{eq:5.38}
\end{eqnarray}
In this alternative form the metric appears exactly once in each
term.

Eqns. (\ref{eq:5.26}), (\ref{eq:5.39}) and (\ref{eq:5.38}) are the
main result of this work. They extend the results of Pletnev and Banin
to curved space-time, and can be used immediately to compute diagonal
matrix elements by means of (\ref{eq:4.5a}). Obvious applications are
the computation of the heat kernel in the non minimal case within a
strict covariant derivative expansion. Such calculation has been
carried out, both for traced and untraced coefficients, and it will be
presented elsewhere. Other interesting application is to the
computation of the effective action of fermions with chiral gauge and
curvature connections. This type of calculation has been done in the
flat space-time case within a covariant derivative expansion for both the
normal and abnormal parity components of the effective action in
\cite{Salcedo:2000hx,Salcedo:2000hp}. So it would seem natural to
extend such results to the case of curved space-time.

\section{Sample computation using covariant symbols}
\label{sec:6}

For the purposes of illustration, in this Section we apply the
method of covariant symbols  to the computation of the diagonal matrix
element of a concrete operator. In Appendix \ref{app:B} we carry out
the analogous computation using the method of symbols.
As operator we take
\begin{equation}
\hat{\cal Q}_{\mu\nu}= \nabla_\mu\frac{1}{m^2-\Delta}\nabla_\nu
\,,
\label{eq:6.1}
\end{equation}
where $m$ is a positive constant c-number. The operator is defined on
a $d$-dimensional Euclidean space-time. $d$ is kept arbitrary so that
ultraviolet convergence of the matrix element is assured in the sense
of dimensional regularization. Note that, through a standard
functional transform, the operator can be related to
\begin{equation}
\hat{\cal H}_{\mu\nu}= \nabla_\mu e^{\tau\Delta}\nabla_\nu
\end{equation}
which is well behaved in the ultraviolet (for positive $\tau$).

The covariant derivative $\nabla_\mu$ includes gauge and world
connections, the latter being the Riemannian connection. We do not
specify the gauge connection and also the space of states is kept
unspecified. In particular, the states may have any tensorial
structure. Also the operator itself is not a world scalar.

We have chosen ${\cal Q}_{\mu\nu}$ instead of $(m^2-\Delta)^{-1}$ or
$e^{\tau\Delta}$ in order to illustrate the method with an operator
that cannot be obtained as a variation of the heat kernel, for which
many results and alternative procedures are available.

Specifically, we will compute
\begin{equation}
{\cal Q}_{\mu\nu}(x)= \langle x|\hat{\cal Q}_{\mu\nu}|x\rangle
\end{equation}
through second derivatives, i.e., neglecting terms with four or more
covariant derivatives. Because $m^2$ is a constant, in this case the
derivative expansion is equivalent to an inverse mass expansion. The
terms neglected introduce a relative error ${\cal O}(1/m^4)$.

Using the relation (\ref{eq:4.5}), we can write
\begin{eqnarray}
{\cal Q}_{\mu\nu}(x) &=&
 \left\langle \overline{{\cal Q}}_{\mu\nu}\right\rangle
\end{eqnarray}
where $\overline{{\cal Q}}_{\mu\nu}$ is the covariant symbol of
$\hat{\cal Q}_{\mu\nu}$, and we have introduced the notation ($X$
representing an arbitrary multiplicative quantity here)
\begin{eqnarray}
 \left\langle X \right\rangle &:=&
\frac{1}{\sqrt{g^{(\xi)}(x)}}\int \frac{d^dp_A}{(2\pi)^d}\,\langle x| X|0\rangle \,,
\qquad
X\in {\cal C}(\underline{\nabla}) 
\,.
\end{eqnarray}
Using the homomorphism property of the covariant symbol implies
\begin{eqnarray}
{\cal Q}_{\mu\nu}(x) &=&
\left\langle \overline{\nabla}_\mu\frac{1}{m^2
-\overline{\Delta}}\overline{\nabla}_\nu
\right\rangle\,.
\end{eqnarray}

Next we proceed to substitute the derivative expansion expressions of
$\overline{\nabla}_\mu$ and $\overline{\Delta}$. A simplification
occurs by noting that $\overline{\nabla}_\lambda$ differs from
$p_\lambda$ only by terms with $\partial^\sigma$. As can be seen in
(\ref{eq:5.39}), $\overline{\nabla}_\mu{}^{(n)}$ for $n\ge 1$ has
$\partial^\sigma$ at the right, or also at the left since these
quantities are antihermitian. Thus
\begin{equation}
\left\langle X
\overline{\nabla}_\lambda \right\rangle = \left\langle X p_\lambda
\right\rangle
\,,\qquad 
\left\langle \overline{\nabla}_\lambda X \right\rangle =
\left\langle p_\lambda X \right\rangle \,.
\end{equation}
This gives
\begin{eqnarray}
{\cal Q}_{\mu\nu}(x) &=&
\left\langle p_\mu
\left[ m^2-p_\lambda p^\lambda 
- \overline{\Delta}{}^{(2)}\right]^{-1} p_\nu
\right\rangle 
+ {\cal O}(\nabla^4)
\nonumber \\
&=&
\left\langle p_\mu
\left[ N + N \left(
\frac{1}{6}\R
+ (Z_{\alpha \beta} 
-\frac{1}{3}\Rc_{\alpha \beta}) \, p^\alpha \partial^\beta
+\frac{1}{3} R_{\alpha \lambda \beta \sigma} 
\, p^\alpha p^\beta \partial^\lambda \partial^\sigma
\right) N \right] p_\nu
\right\rangle 
+ {\cal O}(\nabla^4)
\,,
\end{eqnarray}
where
\begin{equation}
N := \frac{1}{m^2+p^2} \,,\quad p^2 := -p_\lambda p^\lambda \ge 0 \,.
\end{equation}
The momentum derivatives are easily computed using the identities
\begin{eqnarray}
&&[\partial^\alpha,N]= 2 p^\alpha N^2 
\nonumber  \\
&& \langle X \partial^\beta N p_\nu \rangle=
 \langle X (2p^\beta p_\nu N^2+\delta_\nu^\beta N ) \rangle
\nonumber  \\
&& \langle X \partial^\alpha\partial^\beta N p_\nu \rangle=
 \langle X \left( 2(\delta_\nu^\alpha p^\beta+ \delta_\nu^\beta p^\alpha
+g^{\alpha\beta} p_\nu )N^2 + 8 p^\alpha p^\beta p_\nu N^3
\right)
\rangle 
\,.
\end{eqnarray}
(Of course, one can choose to apply $\partial^\mu$ to the left, by
parts.)  In addition we group together the $p$'s and the $N$'s using
\begin{equation}
p_\mu Z_{\alpha\beta}=  
Z_{\alpha\beta} \, p_\mu + R_{\alpha\beta~\mu}^{~~~\lambda}\, p_\lambda
\,,\qquad
[Z_{\mu\nu},N]=0 \,.
\end{equation}
This produces\footnote{Terms with four $p$'s cancel among them. It is
not obvious to me whether this is just accidental or to be expected a
priori.}
\begin{eqnarray}
{\cal Q}_{\mu\nu}(x) &=&
\left\langle p_\mu p_\nu N
+\frac{1}{6} \R \, p_\mu p_\nu N^2
+(Z^\alpha_{~\nu}-\frac{1}{3} \Rc^\alpha_{~\nu}) p_\alpha p_\mu N^2
+R_{~\mu~\nu}^{\alpha ~\beta} \, p_\alpha p_\beta N^2
\right\rangle 
+ {\cal O}(\nabla^4)
\,.
\end{eqnarray}

The momentum integrals can already be taken. In principle, for an
expression $\left\langle X \right\rangle$ ($X$ being a multiplicative
operator) the proper procedure would be to take matrix elements
$\langle x|X|0\rangle$, and then proceed to carry out the integration
over the $d$ constants $p_A$. In the present case, all the $p$
dependence is contained in blocks of the type $(p_{\mu_1}\cdots
p_{\mu_{2r}} N^n)$, where all quantities commute among them (all $p$'s
have been put together in each term). Thus in this case we can
equivalently carry out the momentum integration in each block using
\begin{eqnarray}
\left\langle
p_{\mu_1}\cdots p_{\mu_{2r}} N^n
\right\rangle_p
:=
\frac{1}{\sqrt{g^{(\xi)}(x)}}\int \frac{d^dp_A}{(2\pi)^d}\, p_{\mu_1}\cdots p_{\mu_{2r}} N^n 
=
\frac{1}{\sqrt{g(x)}}\int \frac{d^dp_\mu}{(2\pi)^d}\, p_{\mu_1}\cdots p_{\mu_{2r}} N^n 
\,.
\label{eq:6.13}
\end{eqnarray}
In addition, we can introduce the multiplicative operator $\hat{\cal
Q}^\prime_{\mu\nu}$ such that
\begin{equation}
{\cal Q}_{\mu\nu}(x) = \langle x|\hat{\cal Q}^\prime_{\mu\nu}|0\rangle \,.
\end{equation}
Using these definitions, we obtain
\begin{eqnarray}
\hat{\cal Q}^\prime_{\mu\nu} &=&
\left\langle p_\mu p_\nu N
\right\rangle_p 
+\frac{1}{6} \R \left\langle p_\mu p_\nu N^2\right\rangle_p 
+(Z^\alpha_{~\nu}-\frac{1}{3} \Rc^\alpha_{~\nu}) 
\left\langle
p_\alpha p_\mu N^2
\right\rangle_p 
+ R_{~\mu~\nu}^{\alpha ~\beta} \,\left\langle  p_\alpha p_\beta N^2
\right\rangle_p 
+ {\cal O}(\nabla^4)
\,.
\end{eqnarray}

As shown in Appendix \ref{app:C} the momentum integrals can be
computed to yield formally the same result as in flat space-time
except that the flat metric is replaced by the metric tensor at $x$. It is
often convenient to apply first angular averages, namely,
\begin{eqnarray}
\left\langle
p_\mu p_\nu f(p^2)
\right\rangle_p 
=
\left\langle
- \frac{p^2}{d}g_{\mu\nu} f(p^2)
\right\rangle_p 
\end{eqnarray}
or more generally
\begin{equation}
p_{\mu_1}\cdots p_{\mu_{2n}} \mapsto 
 \frac{(-p^2)^n}{d(d+2)\cdots (d+2n-2)} g_{\mu_1\cdots \mu_{2n}}
\,,
\end{equation}
where $g_{\mu_1\cdots \mu_{2n}}$ is the completely symmetric sum of
$n$-products of metrics ($(2n-1)!!$ terms).

The angular average yields
\begin{eqnarray}
\hat{\cal Q}^\prime_{\mu\nu} &=&
-\frac{1}{d}g_{\mu\nu} \left\langle p^2 N\right\rangle_p 
-\frac{1}{d}
\left(\frac{1}{6} g_{\mu\nu} \R + \frac{2}{3} \Rc_{\mu\nu} + Z_{\mu\nu} \right)
\left\langle p^2 N^2\right\rangle_p
+ {\cal O}(\nabla^4)
\,.
\end{eqnarray}
Finally, the standard formulas of dimensional integration apply,
\begin{eqnarray}
\left\langle (p^2)^r N^n\right\rangle_p
= \frac{(m^2)^{d/2+r-n}}{(4\pi)^{d/2}}\frac{\Gamma(d/2+r)}{\Gamma(d/2)}
\frac{\Gamma(n-d/2-r)}{\Gamma(n)}
\,,
\end{eqnarray}
and so
\begin{eqnarray}
\hat{\cal Q}^\prime_{\mu\nu} &=&
\frac{m^d}{(4\pi)^{d/2}}\Gamma(1-d/2)
\left[
\frac{1}{d}g_{\mu\nu} 
-\frac{1}{m^2} \left(
\frac{1}{12} g_{\mu\nu} \R + \frac{1}{3} \Rc_{\mu\nu} + \frac{1}{2} Z_{\mu\nu} \right)
\right]
+ {\cal O}(\nabla^4)
\,.
\label{eq:6.18}
\end{eqnarray}
This result is manifestly covariant, and formally independent of the
domain and range of $\hat{\cal Q}_{\mu\nu}$, as $\hat{\cal
Q}_{\mu\nu}$ itself in (\ref{eq:6.1}). To fully fix the matrix element it remains to
specify the internal and world structures of the states $\langle x,a,w|$ and
$|0,b,w^\prime\rangle$.

\section{Concluding remarks}
\label{sec:7}

From the computational point of view, our main result is contained in
Eqns. (\ref{eq:5.26}), (\ref{eq:4.14}) (for a general connection) and
(\ref{eq:5.39}) and (\ref{eq:5.38}) for the Riemannian
connection. With such building blocks, and with the help of the
representation (homomorphism) property, one can construct the
covariant symbol of other operators $f(\nabla,M)$. This has been
illustrated with an explicit computation in Section \ref{sec:6}. It
clearly would be interesting to extend the present results to higher
orders and to include torsion more systematically. See e.g.
\cite{Nieh:1981xk,Chandia:1997hu} for relations between torsion and
chirality. For references motivating the study of quantum field theory
in curved space-time with torsion see
\cite{Buchbinder:1985ux,Buchbinder:1990ku,Buchbinder:1985ym}.

Regarding the concrete expressions obtained, we observe that they are
rather natural, involving local covariant operators, similar to the
heat kernel coefficients.\footnote{It is noteworthy that actually not
all such operators are present and thus selection rules are at
work. For instance, terms of the form $M_{ss}\Rc_{ss}(\partial^s)^4$
do not appear in $\overline{M}$. Technically, the reason for the non
existence of such terms is that $\overline{M}$ does not involve
$p_\mu$ (cf. (\ref{eq:5.12})). Then, using only $\partial^\mu$ and
derivatives of $M$, $Z$'s, the Riemann tensor and the torsion, there
is no way to contract all indices. The presence of $p_\mu$ would
permit $[\partial^\nu,p_\mu]=\delta^\nu_\mu$, and hence the Ricci
tensor to appear. This is an alternative proof of
(\ref{eq:5.26}).\label{foot:19}} However, we warn that the presence of
$Z_{\mu\nu}$ (or other operators in ${\cal C}(Z)$) is unusual as
compared to other treatments. In those treatments
\cite{Vassilevich:2003xt}, if one needs to apply, say, the heat kernel
operator on a state with world indices (e.g., the gluon field,
$G_\mu$), a first step is to transform the world index into an
internal one using a tetrad field, $G_a=e^\mu_a G_\mu$. The new field
$G_a$ is a coordinate scalar so one can apply the heat kernel
expressions for scalars. The world structure of the field is now in
the internal sector through the corresponding connection for the
tetrad index. In this way the result depends on an $F_{\mu\nu}(x)$
which includes the strength tensor from the original gauge structure
plus that of the new internal structure. The idea is to assimilate
coordinate covariance as much as possible to the gauge case, where
$F_{\mu\nu}(x)$ is a matrix valued function. Our own representation is
different since the expressions obtained in the present work hold
regardless of the world tensor structure of the states, without
transforming them into scalars. This works thanks to the action of
$Z_{\mu_1\ldots\mu_n}$ which are not just matrix valued fields: from
the gauge point of view, while $F_{\mu\nu}$ is the same matrix valued
function in, e.g. $F_{\mu\nu}B_\alpha|~\rangle$ and in $B_\alpha
F_{\mu\nu}|~\rangle$, $Z_{\mu\nu}$ would be a ``different'' matrix
valued function in each case, since it acts on any world index at its
right. Of course, nothing prevents us to reduce our formulas to
reproduce the abovementioned more usual point of view. To do so, once
the state has been transformed into a world scalar, one only needs to
move all $Z_{\mu\nu}$ to the right using commutators, and then set the
world part of $Z_{\mu\nu}$ to zero (since it is acting on a world
scalar).  Nevertheless, in our view, it is more natural to work with
the original fields rather than transforming them into scalars by
means of an ad hoc new internal structure.

With respect to applications of the method exposed, it naturally
applies to one loop computations is curved space-time. A first
application would be to compute the heat kernel, not using the
standard Seeley-DeWitt expansion, which orders operator by their
dimension, but the covariant derivative expansion. Explicit
calculations along this line exist only for the minimal case (i.e.,
Klein-Gordon theories with a trivial gauge sector)
\cite{Gusynin:1990bu,Moss:1999wq}. The non-minimal, but flat
space-time, calculation of \cite{Salcedo:2004yh} can be extended to
the curved case and results for traced and untraced coefficients will
be presented elsewhere. Further applications refer to the effective
action of Klein-Gordon and Dirac theories in curved space-time. Again
results obtained by the covariant symbol method exist for these two
cases, for flat space-time and quite general non Abelian backgrounds
\cite{Salcedo:2000hp,Salcedo:2000hx}.  These computations correctly
reproduce the Wess-Zumino-Witten action
\cite{Wess:1971yu,Witten:1983tw} as well as the associated anomalies
in the abnormal parity sector of the fermion case. For Dirac fermions
in curved space-time there are many interesting results concerning
chiral, coordinate and frame anomalies
\cite{Alvarez-Gaume:1983ig,Bardeen:1984pm,%
Leutwyler:1985ar,Alvarez-Gaume:1985dr,Bertlmann:1996bk} but results
are much more scarce for the effective action itself. We expect that
all the anomalies will be obtained as a byproduct of the effective
action computation.

Mathematically, the covariant symbol is a quite interesting and
challenging quantity, since it implies the construction of a true
representation of pseudodifferential operators in terms of purely
multiplicative operators (in the original $x$ space). It would be very
nice to have any rigorous result concerning such quantities and in
particular to obtain the exact covariant symbol in particular
cases. The fact that the covariant symbol can be computed
systematically within concrete (presumably asymptotic) expansions
suggests that this quantity can be given a rigorous and proper
mathematical definition.

\begin{acknowledgments}
This work was supported by DGI, FEDER, UE and Junta de Andaluc{\'\i}a funds 
(FIS2005-00810, HPRN-CT-2002-00311, FQM225).
\end{acknowledgments}

\appendix

\section{Notational conventions}
\label{app:A}

In this appendix we summarize the non standard notational conventions used in the main text.

\medskip
{\bf Derivative convention.} For a quantity $X_I$ having an ordered
set of world indices $I$, $X_{\mu I}$ denotes its covariant derivative
$[\nabla_\mu,X_I]$. E.g.
$R_{\sigma\mu\nu~\beta}^{~~~\,\,\alpha}:=[\nabla_\sigma,
R_{\mu\nu~\beta}^{~~\,\alpha}]$. Exceptions are
$Z_{\mu_1\mu_2\cdots\mu_n}$, defined in (\ref{eq:Zs}) so that they are
multiplicative operators with respect to $x$, and the tensors
$P_{\mu_1\mu_2\cdots\mu_n}\!{}^\alpha_{~\,\beta}$ defined in
(\ref{eq:4.13}).

Note that on states (wavefunctions) the action of $\nabla_\mu$ is
expressed as $\nabla_\mu \psi$ while on operators it acts adjointly,
$[\nabla_\mu,X]$. Occasionally will write simply $\nabla_\mu X$ if
$X \in{\cal
C}(\underline{\nabla},\underline{Z},\underline{I})$), e.g.
$\nabla_\sigma R_{\mu\nu~\beta}^{~~\,\alpha}$ or
$\nabla_\mu t^A_\nu$.

\medskip
{\bf Momentum convention.} The momentum $p_\mu$ is purely imaginary,
$p_\mu=ik_\mu$ ($k_\mu$ real). However, $d^dp$ is just the standard
real measure $d^nk$ and $p^2:=-g^{\mu\nu}p_\mu p_\nu= g^{\mu\nu}k_\mu
k_\nu$ the standard real norm (positive for Euclidean signature).

\medskip
{\bf $s$ index convention.} $s$ indicates a symmetrized world
index. So we will use the notation
\begin{equation}
A_{s\mu ss}(\partial^s)^3
\end{equation}
to mean
\begin{equation}
A_{\alpha\mu\beta\gamma}\partial^\alpha \partial^\beta \partial^\gamma
\,.
\end{equation}
There is no ambiguity since the $\partial^\mu$ commute.\footnote{Note
that in an expression like $\partial^\alpha
Z_{\alpha\beta\mu}\partial^\beta$ the indices $\alpha$ and $\beta$ are
not symmetrized ($Z$ and $\partial$ do not commute). The expression
differs from $\partial^\beta Z_{\alpha\beta\mu}\partial^\alpha$ and
hence it would not be faithfully represented by $\partial^s
Z_{ss\mu}\partial^s$.} In section \ref{sec:5} we use a similar
convention for the index $S$, which refers to the labels of the type
$A$, $B$, etc.

\medskip
{\bf Riemann tensor convention.} For a gauge singlet and world vector wavefunction
\begin{equation}
[\nabla_\mu,\nabla_\nu] V^\lambda= +R_{\mu\nu~\,\sigma}^{~~\,\,\lambda~}V^\sigma
-T_{\mu\nu}^{~~\sigma}\,\nabla_\sigma V^\lambda \,,\quad
\end{equation}
The Ricci tensor and the scalar curvature are
\begin{equation}
\Rc_{\mu\nu}:=R_{\lambda\mu~\nu}^{~~\,\,\lambda~}
\,,\quad
\R:=g^{\mu\nu}\Rc_{\mu\nu} \,.
\end{equation}

\section{Computation of ${\cal Q}_{\mu\nu}(x)$ using the method of symbols}
\label{app:B}

In this appendix we will illustrate the method of (ordinary) symbols
with the same operator $\hat{\cal Q}_{\mu\nu}$ considered in Section
\ref{sec:6}. (We will use definitions introduced in that Section.)
Eq. (\ref{eq:3.2d}) implies
\begin{eqnarray}
{\cal Q}_{\mu\nu}(x) &=& 
\frac{1}{\sqrt{g^{(\xi)}(x)}}
\int \frac{d^dp_A}{(2\pi)^d}\langle x|{\cal Q}_{\mu\nu}(\nabla+p,M)|0\rangle 
\nonumber \\
&=&
 \left\langle
 (\nabla_\mu+p_\mu)\frac{1}{m^2-(\nabla_\alpha+p_\alpha)(\nabla^\alpha+p^\alpha)}
(\nabla_\nu+p_\nu)
\right\rangle .
\end{eqnarray}

Carrying out an expansion in powers of $\nabla_\mu$ through second order yields
\begin{eqnarray}
{\cal Q}_{\mu\nu}(x) &=& 
 \Big\langle
p_\mu p_\nu N 
+ p_\mu N \Big(
\Delta +\{\nabla_\alpha,p^\alpha \}N \{\nabla_\beta, p^\beta \}
\Big) N p_\nu
\nonumber \\
&&
+p_\mu N \{\nabla_\alpha,p^\alpha \} N \nabla_\nu
+ \nabla_\mu N \{\nabla_\alpha,p^\alpha \} N  p_\nu
+ \nabla_\mu N \nabla_\nu
\Big\rangle
+ {\cal O}(\nabla^4)
\,.
\end{eqnarray}
The next step is to move all $\nabla$'s to (say) the right (it is
essential not to split the covariant derivative into non covariant
pieces). The move is obtained by applying the rules
\begin{equation}
[\nabla_\mu,p_{\alpha_1\cdots\alpha_n}]= p_{\mu\alpha_1\cdots\alpha_n} \,,
\quad
[\nabla_\mu,N] = 2 p^\alpha p_{\mu\alpha} N^2
\,,
\end{equation}
this gives
\begin{eqnarray}
{\cal Q}_{\mu\nu}(x) &=& 
 \Big\langle
 p_\mu p_\nu N
+  p_\mu p^\alpha_{~\alpha\nu} N^2 
+ p_\nu p^{~\,\alpha}_{\mu ~\alpha} N^2 
+ 2  p^\alpha p_{\mu\alpha\nu} N^2
+ 2  p_\mu p_\nu p^\alpha p^{~\,\beta}_{\alpha~\beta} N^3
\nonumber \\ &&
+ 2  p_\mu p_\nu p^\alpha p^\beta_{~\beta\alpha} N^3
+ 4  p_\mu p^\alpha p^\beta  p_{\alpha\beta\nu} N^3
+ 4  p_\nu p^\alpha p^\beta  p_{\mu\alpha\beta} N^3
+ 8  p_\mu p_\nu p^\alpha p^\beta p^\gamma  p_{\alpha\beta\gamma} N^4
\nonumber \\ &&
+ N \nabla_\mu \nabla_\nu 
+   p_\mu p_\nu N^2 \nabla^\alpha \nabla_\alpha 
+ 2  p_\mu p^\alpha N^2 \nabla_\alpha \nabla_\nu
\nonumber \\ &&
+ 2 p_\nu p^\alpha N^2 \nabla_\mu \nabla_\alpha
+ 4 p_\mu p_\nu p^\alpha p^\beta N^3 \nabla_\alpha \nabla_\beta
%
+\text{~39~further~terms~}
\Big\rangle
+ {\cal O}(\nabla^4)
\,.
\end{eqnarray}
The ``39 further terms'' not made explicit contain a factor
$p_{\alpha\beta}$. For the Riemannian connection they vanish by
choosing normal coordinates centered at $x$. (Of course, the rule
$p_{\alpha\beta}=0$ can only be applied after all covariant
derivatives have been put aside.)

The derivatives of $p_\mu$ are easily obtained  recalling that
$p_\alpha= t^A_\alpha p_A$, thus
\begin{equation}
p_{\mu_1\cdots \mu_n}=
t^A_{\mu_1\cdots \mu_n} p_A=
t^A_{\mu_1\cdots \mu_n} t_A^\lambda \, p_\lambda \,.
\end{equation}
In particular, using (\ref{eq:5.22b}) one obtains
\begin{equation}
p_{\alpha\mu\nu}=
\frac{1}{3}\left(
R^\lambda_{~\mu\nu\alpha}+ R^\lambda_{~\nu\mu\alpha}\right)
 p_\lambda \,.
\end{equation}
Substitution in the expression above gives
\begin{eqnarray}
{\cal Q}_{\mu\nu}(x) &=& 
 \Big\langle
 p_\mu p_\nu N
-\frac{2}{3} \Rc_{\mu\alpha} \, p_\nu p^\alpha N^2
+\frac{1}{3} \Rc_{\nu\alpha} \, p_\mu p^\alpha N^2
+\frac{2}{3} R_{\mu\alpha\nu\beta} \, p^\alpha p^\beta N^2
-\frac{2}{3} \Rc_{\alpha\beta} \, p_\mu p_\nu p^\alpha p^\beta N^3
\nonumber \\ &&
+ N \nabla_\mu \nabla_\nu 
+   p_\mu p_\nu N^2 \nabla^\alpha \nabla_\alpha 
+ 2  p_\mu p^\alpha N^2 \nabla_\alpha \nabla_\nu
\nonumber \\ &&
+ 2 p_\nu p^\alpha N^2 \nabla_\mu \nabla_\alpha
+ 4 p_\mu p_\nu p^\alpha p^\beta N^3 \nabla_\alpha \nabla_\beta
\Big\rangle
+ {\cal O}(\nabla^4)
\,.
\end{eqnarray}
Prior to momentum integration, this expression is not manifestly
covariant since there are still $\nabla$'s not derivating anything nor
in the form of $Z_{\mu_1\cdots\mu_n}$, that is, the expression in
brackets is not a multiplicative operator. It is often not necessary to
completely carry out the momentum integration to achieve manifest
gauge covariance \cite{Chan:1986jq,Caro:1993fs}. Taking an angular average, as
explained in Section \ref{sec:6}, gives
\begin{eqnarray}
{\cal Q}_{\mu\nu}(x) &=& 
 \Big\langle
-\frac{p^2}{d}g_{\mu\nu} N
-\frac{1}{3} \frac{p^2}{d}\Rc_{\mu\nu} N^2
- \frac{2}{3}\frac{(p^2)^2}{d(d+2)} g_{\mu\nu} \R N^3
- \frac{4}{3}\frac{(p^2)^2}{d(d+2)} \Rc_{\mu\nu} N^3
\nonumber \\ &&
+ N \nabla_\mu \nabla_\nu 
- 4  \frac{p^2}{d}  N^2 \nabla_\mu \nabla_\nu
- \frac{p^2}{d} g_{\mu\nu} N^2 \nabla^\alpha \nabla_\alpha 
\nonumber \\ &&
+ 4  \frac{(p^2)^2}{d(d+2)}  N^3 \nabla_\mu \nabla_\nu
+ 4  \frac{(p^2)^2}{d(d+2)}  N^3 \nabla_\nu \nabla_\mu
+ 4\frac{(p^2)^2}{d(d+2)} g_{\mu\nu}  N^3 \nabla^\alpha \nabla_\alpha 
\Big\rangle
+ {\cal O}(\nabla^4)
\,.
\end{eqnarray}
Using now the recurrence
\begin{eqnarray}
\left\langle (p^2)^r N^n\right\rangle_p
= \frac{d/2+r-1}{n-1} \left\langle (p^2)^{r-1} N^{n-1}\right\rangle_p
\,, 
\qquad n > 1
\end{eqnarray}
to eliminate higher powers of $N$, gives already a covariant result
\begin{eqnarray}
{\cal Q}_{\mu\nu}(x) &=& 
 \Big\langle
-\frac{p^2}{d}g_{\mu\nu} N
- \frac{1}{12} g_{\mu\nu} \R N
- \frac{1}{3}\Rc_{\mu\nu} N
%
- \frac{1}{2}  N [\nabla_\mu, \nabla_\nu]
\Big\rangle
+ {\cal O}(\nabla^4) \,.
\end{eqnarray}
Upon momentum integration this coincides with (\ref{eq:6.18})
obtained in Section \ref{sec:6} using the method of covariant symbols.

\section{Momentum integrals}
\label{app:C}

Here we want to show that momentum integrals like (\ref{eq:6.13}) give
formally the same result as in flat space-time. Rather than
considering the most general case, it will be sufficient to treat a
sample integral. We consider
\begin{equation}
I_{\mu\nu} =  \frac{1}{\sqrt{g}}\int \frac{d^dk}{(2\pi)^d}\, k_\mu k_\nu f(k^2) \,.
\end{equation}
Here $k_\mu$ are real ($p_\mu :=i k_\mu$, $d^dp := d^dk$, cf. Appendix
\ref{app:A}) and $k^2 = g^{\alpha\beta}k_\alpha k_\beta$

Introducing an orthonormal tetrad, $e^a_\alpha$, as well as a new
momentum variable, $q_a$,
\begin{equation}
e^\alpha_a e^a_\beta = \delta^\alpha_\beta \,,\qquad
g_{\alpha\beta}= \eta_{ab}e^a_\alpha e^b_\beta 
\,,\qquad k_\alpha= e^a_\alpha q_a 
\end{equation}
we obtain
\begin{equation}
g= |\det g_{\alpha\beta}|= |\det e^a_\alpha|^2
\,,\qquad
d^dk= |\det e^a_\mu| \,d^dq= \sqrt{g} \, d^dq
\,,\qquad
k^2= \eta^{ab}q_a q_b = q^2
\end{equation}
 so
\begin{equation}
I_{\mu\nu} =  I_{ab}\,e^a_\mu e^b_\nu
\end{equation}
with
\begin{equation}
I_{ab}= \int \frac{d^dq}{(2\pi)^d}\, q_a q_b f(q^2) = \eta_{ab}I \,,
\end{equation}
and finally
\begin{equation}
I_{\mu\nu} =  g_{\mu\nu}I \,,
\end{equation}
with $I$  computed as in flat space-time
\begin{equation}
I= \int \frac{d^dq}{(2\pi)^d}\, \frac{1}{d}q^2 f(q^2)\,.
\end{equation}


\end{document}